\documentclass[fleqn,usenatbib]{mnras}

\usepackage{newtxtext,newtxmath}
\usepackage[T1]{fontenc}
\usepackage{ae,aecompl}

\usepackage{graphicx}	% Including figure files
\usepackage{amsmath}	% Advanced maths commands
\usepackage{amssymb}	% Extra maths symbols
\usepackage{longtable}
\usepackage{lscape}
\usepackage[utf8]{inputenc}
\usepackage{supertabular}
\usepackage{graphicx}

\newcommand{\be}{\begin{equation}}
\newcommand{\en}{\end{equation}}

\def\lya{Ly$\alpha$}

\def\mgii{Mg~{\sc ii}}

\def\ni2{Ni~{\sc ii}}
\def\nv{N~{\sc v}}

\def\alii{Al~{\sc ii}}
\def\aliii{Al~{\sc iii}}
\def\civ{C~{\sc iv}}
\def\siiv{Si~{\sc iv}}

\def\si2{Si~{\sc ii}}
\def\feii{Fe~{\sc ii}}
\def\h1{H~{\sc i}}
\def\feiii{Fe~{\sc iii}}

\def\kms{km s$^{-1}$}
\title[Stripe 82 absorption line variability]{Ionization driven intrinsic absorption line variability of BAL quasars in the Stripe 82 region}

\author[M. Vivek]{
M. Vivek$^{1}$\thanks{E-mail: vzm83@psu.edu; \hspace{0.2cm}getkeviv@gmail.com}\\ 
$^{1}$ 525, Davey Lab, Pennsylvania State University, State College, PA-16802, USA}

\date{Accepted XXX. Received YYY; in original form ZZZ}

\pubyear{2019}

\begin{document}
\label{firstpage}
\pagerange{\pageref{firstpage}--\pageref{lastpage}}
\maketitle

\begin{abstract}
We investigate the connection between the intrinsic \civ\ absorption line variability and the continuum flux changes of  broad absorption line (BAL) quasars using a sample of 78 sources in the Stripe 82 region. The absorption trough variability parameters are measured using the archival multi-epoch spectroscopic data from the Sloan Digital Sky Survey (SDSS), and the continuum flux variability parameters are estimated from the photometric light curves obtained by the SDSS and the Catalina Real-Time Survey (CRTS) surveys.  We find evidence for weak correlations ($\rho_s \sim$ 0.3) between the intrinsic \civ\ absorption line variability and the quasar continuum variability for the final sample of 78 BAL quasars. The correlation strengths improve ($\rho_s \sim$ 0.5) for the ``high-SNR" sample sources that have higher spectral signal-to-noise ratio. Using two sub-sets of the high-SNR sample differing on the absorption trough depth, we find that  the shallow trough sub-set shows an even stronger correlation ($\rho_s \sim$ 0.6), whereas the deep trough sub-set does not show any correlation between the absorption line variability and the continuum variability.  These results point to the important role of saturation effects  in the correlation between the  absorption line variability and the continuum variability of BAL quasars.    Considering other effects that can also smear the correlation, we conclude that the actual correlation between the absorption line and continuum variability is even stronger.  
\end{abstract}


%%%%%%%%%%%%%%%%%%%%%%%%%%%%%%%%%%%%%%%%%%%%%%%%%%

%%%%%%%%%%%%%%%%% BODY OF PAPER %%%%%%%%%%%%%%%%%%

\section{Introduction}
 The well-known correlation in active galactic nuclei (AGN) between the black hole properties and the host-galaxy  properties suggest an intimate coupling between the evolution of the black hole and its host-galaxy \citep[][and references therein]{ferrarese00,gultekin09,shen15}. The process by which this coupling occurs is known as AGN feedback. { Together with the large scale molecular and ionized outflows, outflowing gas originating in the nuclear region of the quasar and detected as blue-shifted broad or narrow absorption lines in the rest-frame UV spectra of quasars, are believed to be the agents that facilitate  AGN feedback \citep[][and references therein]{silk98,dimatteo05,hopkins09}. Although the connection between the different phases of outflows are not clearly understood, many previous studies have shown that nuclear quasar outflows have a significant role in AGN feedback \citep[for eg.,][]{arav13,borguet13a,chamberlain15,cicone18}}. Besides influencing the evolution of the black hole and the host-galaxy, these outflows may be responsible for (a) quenching the star formation of the host-galaxy (b) limiting the growth of the black hole and, (c) transporting heavy elements to the intergalactic medium scales.  Therefore,  understanding the properties of these outflowing gas is important to understand the working of the quasar central-engine as well as the host-galaxy.

Intrinsic absorption lines, produced by gas directly associated with the AGN, are classified into broad, mini-broad and narrow absorption lines (BALs, mini-BALs, and NALs) according to the width of the absorption line. { BALs are defined as lines with a full width at half maximum (FWHM) exceeding 2000 \kms. On the other hand, mini-BALs have  FWHMs of 500--2000 \kms\  and NALs have FWHMs $\leq$ 500 \kms.}  BAL quasars  are further classified into three sub-types based on the ionization state of the absorbing gas. High-ionization BAL quasars (HiBAL quasars) are characterized by absorption from \civ\, \siiv, \nv, and \lya. In addition to the high-ionization lines, low-ionization BAL quasars (LoBAL quasars) also feature absorption from \mgii, \alii\ and \aliii. LoBAL quasars containing excited fine-structure levels of \feii\ or \feiii\ are called iron LoBAL quasars (FeLoBAL quasars). The relative fraction of HiBAL, LoBAL, and FeLoBAL quasars among the total quasar population are 15, 1 and 0.1 percent respectively. The observed  fraction of BAL quasars among the quasar population  (20 percent for optically selected quasars \citep{knigge08} and 40 percent when corrected for selection effects \citep{allen11}) is attributed to either an orientation effect \citep[for eg.,][]{elvis12}, where the observer's line of sight passes through the BAL clouds or an evolutionary phase in the quasar lifetime \citep[for eg.,][]{farrah07}. It has been suggested that mini-BALs are an intermediate stage in BAL structure and/or evolution and the two classes form a continuum of absorber properties \citep{ganguly08,gibson09}. 

Absorption line variability study is an important technique which can constrain the kinematics and physical conditions of the absorbing gas. Previous BAL variability studies have established that BALs are variable on time-scales ranging from a few days \citep[for eg.,][]{grier15,hemler19} to a few years \citep{lundgren07,gibson08,gibson10,capellupo11,capellupo12,filiz12,capellupo13,filiz13,vivek14,welling14,vivek16,mcgraw17,decicco18,rogerson18}. However, the exact cause of  the absorption line variability still remain unclear.   Two mostly favoured mechanisms for the cause of  quasar absorption line variability are (a) gas clouds moving across the line of sight and, (b) changes in optical depth of the absorbers due to changes in the ionizing flux. In the former case, the variability time-scale is useful to set upper limits on the transverse velocity, size, and location of the absorber and in the latter case, the variability time-scale can put  constraints on the density, size, and location of the absorber. { Besides the above two mechanisms, quasar absorption lines can also vary due to changes in the density of the absorber.} Understanding the main driver of absorption line variability in quasars is important for the use of absorption line variability to obtain constraints on the outflow properties. 

{ In the ionization driven absorption line variability scenario, the changes in the ionizing flux can be due to the (a) fluctuations in the quasar radiation field,  or, (b)  changes in the ``shielding gas" located at the base of the outflow.} In radiatively driven accretion disc wind models, the proposed ``shielding gas" acts as a filter to the hard-ionizing radiation and prevents the BAL absorbing cloud from over-ionization  \citep{murray95,proga00}. Previous BAL variability studies have presented evidence in favor of the ionization driven BAL variability \citep[for eg.,][]{barlow94, filiz13, wang15,he17}, absorber motions across the line of sight \citep[for eg.,][]{gibson08,hamann08,hall11,filiz12,vivek12,vivek16} or a combination of both effects \citep[for eg.,][]{capellupo12}. In these studies, coordinated variations of different velocity absorption components of the same ion are attributed to changes in the ionization.  { BAL variabilities occurring over small portions of the troughs are considered  either to be due to clouds crossing the line of sight or due to the difference in the density or the covering fraction of the absorbers at different velocities}. Recently, two studies have reported that BAL variability is mostly caused by changes in the ionization state of the absorbing gas \citep{wang15,he17}.  Using a large sample of Sloan Digital Sky Survey (SDSS)-Data  Release (DR)10  selected quasars, \citet{wang15} reported coordinated variations between absorption lines and continuum/emission lines and claimed that absorption line variability is mainly driven  by changes in the gas ionization in response to continuum variations.  \citet{he17} presented a statistical analysis of a large sample of 843 quasars and showed that BAL variabilities in at least 80 percent quasars are due to the variation of ionizing continuum. 

In this paper, we take advantage of the multi-epoch  SDSS archival data available in the Stripe 82 region of the sky. Stripe 82 region has been repeatedly monitored both in photometry and spectroscopy by several  SDSS programs. We make use of this rich data set to explore the correlations between the intrinsic \civ\ absorption line variability and the continuum variability in quasars. Absorption line variabilities are estimated from the multi-epoch SDSS spectroscopic data, and the continuum variabilities are computed using the well sampled ($\sim$ 60 epochs for each source) photometric measurements from SDSS and Catalina Real-Time Sky (CRTS) survey. Section 2 outlines the spectroscopic and photometric observations and procedures used for constructing the sample. Details regarding measurements of the absorption line and the continuum variabilities are discussed in section 3. Statistical analysis of the measured variabilities is presented in section 4. In section 5, we discuss our main findings in the context of other BAL variability results. 

\begin{figure*}
    \centering
    \includegraphics[scale=0.4]{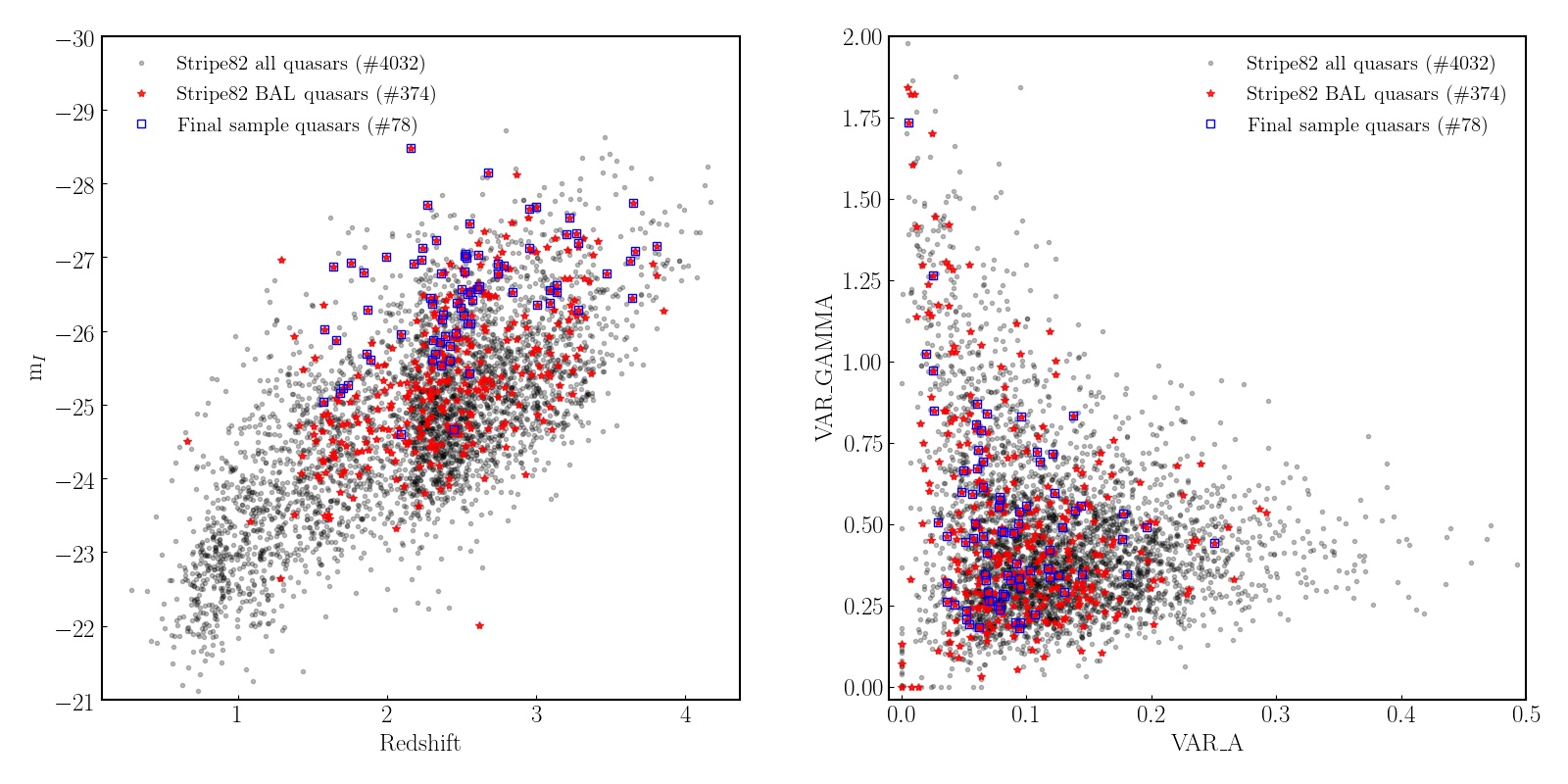}
    \caption{ The distribution of Stripe 82 quasars (black dots), Stripe 82 BAL quasars (red stars) and the sources in the final sample (blue squares) in the redshift-luminosity (left panel) and VAR\_A-VAR\_GAMMA (right panel) plane.}
    \label{fig:z_MI_plot}
\end{figure*} 
\section{Sample selection and observations}
{ Photometric variability of quasars are often parametrized by the ``structure function" which is a measure of the amplitude of the variability as a function of the time delay between the observations \citep[for e.g.,][]{giveon99,chelsea10}.
\begin{equation}
    V(\Delta t) = \left\langle \sqrt{\frac{\pi}{2}} |\Delta m_{ij}| - \sqrt{\sigma_i^2+\sigma_j^2} \right\rangle_{\Delta t}
\end{equation}
where $\Delta m_{ij}$ is the magnitude difference at time t$_i$ and t$_j$ and $\sigma_i$, $\sigma_j$ are the  photometric errors at time t$_i$ and t$_j$.

Assuming a Gaussian distribution for the noise and the intrinsic photometric variability, the structure function is modeled as a power-law parametrized in terms of A, the rms magnitude difference on a one year and $\gamma$,  the logarithmic gradient of the mean change
in magnitude with time \citep[e.g.,][]{schmidt10}.
\begin{equation}
    V(\Delta t|A,\gamma) = A\left( \frac{\Delta t}{1 yr} \right)^{\gamma}
\end{equation}
Large values of A indicate large fluctuation amplitudes and large values of $\gamma$ indicate an increase of the fluctuation amplitude with time \citep[See][for a detailed discussion on Stripe 82 quasar variability]{nathalie11}.} %The model predicts an evolution of the variance $\sigma^2$($\Delta$m) with time according to,
 %       \begin{equation}
 %           \sigma^2(\Delta m_{ij}) = [A(\Delta t_{ij})^\gamma]^2 + (\sigma_i^2+\sigma_j^2)
 %       \end{equation}

Our initial quasar sample is mainly drawn from the SDSS DR12 Quasar (hereafter, DR12Q)  catalog of \citet{paris17}. We used the DR12Q catalog rather than the latest DR14Q catalog as the DR12Q catalog also includes estimates of quasar  structure function parameters. DR12Q catalog contains information about the structure function parameters, A and $\gamma$, in the form of keywords VAR\_A and VAR\_GAMMA respectively { (hereafter, we use VAR\_A and VAR\_GAMMA to designate A and $\gamma$ parameters obtained from the DR12Q catalog)}.  We first selected the sample to only include quasars which have at least  $>$ 10 epochs of photometric observations (i.e., DR12Q catalog keyword VAR\_MATCHED $>$ 10). 
This selection ensures that all sources in our sample have a good number of photometric observations and the continuum variability estimates are reliable. The quasars in the Stripe 82 region each have $\sim$ 60  epochs of photometric observations and the DR12Q structure function parameters for this Stripe 82 quasar sample were estimated using the SDSS photometric observations alone.  For the remaining quasars in the SDSS footprint, the structure function parameters were estimated using 3-10 photometric epoch data from SDSS and Palomer Transient Factory \citep[PTF;][]{rau09} observations.   We notice  that a small  fraction of quasars in the parent DR12Q sample have non-physical  A, $\gamma$ values   even for  sources with  more than 10 photometric epochs. { The spurious A-$\gamma$ values in the parent DR12Q sample may be associated either with the systematics  arising from the use of two different sources of photometric data (for eg., PTF magnitudes correspond to coadded imaging data from a few tens to thousands  of single-epoch PTF images) or to the relatively lower number of photometric epochs.}  As all the quasars in the Stripe 82 region  have superior light curve information,  we  limited our sample to only include sources in the Stripe 82 region. We further reduced our sample size to only include BAL quasars with at least two epochs of spectroscopic data from SDSS-I/II \citep[][]{york00} and SDSS-III \citep[][]{eisenstein11,dawson13} surveys. As we are mainly interested in studying the \civ\ absorption lines, we eliminated sources that have redshifts below 1.5. To select sources that have the two spectroscopic data taken on sufficiently separated dates, we imposed a new condition that the difference between the spectroscopic epochs should be greater than 1000 days in the observer time-scale. { The resulting sample of 78 sources forms the main sample of our analysis. Hereafter, we refer to this sample of 78 sources as the ``final sample".} Fig.~\ref{fig:z_MI_plot} shows the distribution of Stripe 82 quasars, Stripe 82 BAL quasars and the final sample sources in the redshift-luminosity plane (left panel) and VAR\_A-VAR\_GAMMA plane (right panel). It is not surprising to note that the final sample sources, which also have spectroscopic data, are relatively brighter as compared to the whole Stripe 82 quasar sample. We further noticed that the signal-to-noise ratios of the SDSS-I/II spectroscopic observations for a large number of sources in the final sample are low. \citet{gibson09b} defined the SN$_{1700}$ as the median of the ratio of flux to noise per pixel over the rest-frame region from 1650-1750 \AA. {  To mitigate the effects of low signal-to-noise ratio (SNR) spectra, we also constructed a ``high-SNR sample" of 45 sources by imposing a cut that  SN$_{1700 \AA}$ > 4 for all spectroscopic epochs. }
Table~\ref{tab:table_summary_selection} summarizes the sample selection procedure implemented in this paper. 

\begin{figure}
    \centering
    \includegraphics[scale=0.37]{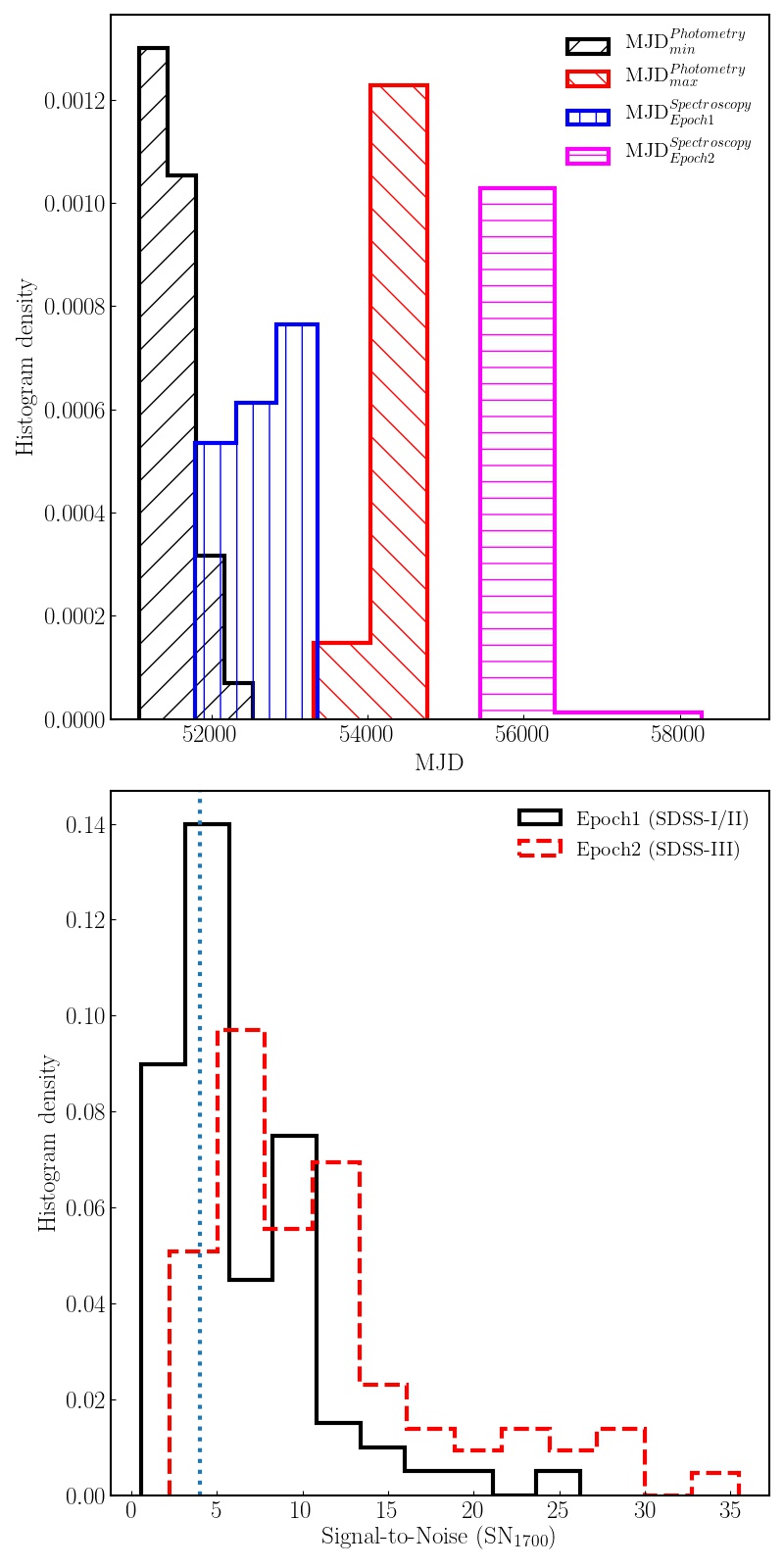}
    \caption{Top Panel: Histograms showing the distribution of the minimum and the maximum MJDs of the photometric observations and the distribution of MJDs of the SDSS-I/II and  SDSS-III spectroscopic observations.  Note that the histogram of the MJDs of the SDSS-I/II spectroscopic observation  is distributed between the histograms of minimum MJDs and maximum MJDs of photometric observations.  Similarly, the histogram of the MJDs of the SDSS-III spectroscopic observation is distributed after the histogram of maximum MJDs of photometric observations. Bottom Panel: Distribution of the signal-to-noise ratio (SN$_{1700}$) for the SDSS-I/II and SDSS-III spectra in the final sample. The dashed/blue vertical line denotes the cut, SN$_{1700}$ = 4, for selecting the high-SNR sample. }
    \label{fig:z_MJDdiff_plot}
\end{figure}
\begin{table}
    \centering
    \begin{tabular}{clr}
    \hline
    (1)& DR12 quasars with variability information     & 143359  \\
    (2)& (1)  and N$_{\mathrm{epoch}}$ (light curve) $ >$ 10     &  10957\\
    (3)& (2) and quasar in the Stripe 82 region & 4032 \\
    (4)& (3) and quasar = BAL quasar & 374 \\
    (5)& (4) and N$_{\mathrm{spectra}} >$ 1 & 90 \\
    (6)& (5) and z $>$ 1.5 & 87 \\
    \hline
    (7)& (6)  and $\Delta$MJD$^{\mathrm{spectra}} >$ 1000 (final sample) & 78 \\
    (8)& (7) and SN$_{1700}^{\mathrm{spectra}} >$ 4 (high-SNR sample) & 45 \\
    \hline
    \end{tabular}
    \caption{Summary table describing the selection of the final sample and high-SNR sample.}
    \label{tab:table_summary_selection}
\end{table}

\subsection{Spectroscopic observations}
The spectra were obtained as part of the SDSS-I/II  and SDSS-III  surveys.  SDSS-I/II data were obtained between the years 2000 and 2009 using a spectrograph that has a wavelength coverage of \mbox{$\sim$3800--9200}~\AA, have pixels 70~km~s$^{-1}$ wide and have resolutions ranging from \mbox{$\sim$1850} to 2200. SDSS-III survey operated between the years 2009 and 2014 and the data were  acquired using an improved spectrograph which cover wavelengths from \mbox{$\sim$3600} to 10\,400~\AA, have pixels 70~km~s$^{-1}$ wide, and have resolutions between \mbox{$\sim$1300--3000} \citep[see][ for details regarding  both the spectrographs]{smee13}.  Among the 78 sources in the final sample, 56, 11, 8 and 3 sources have two, three, four and five epochs of spectroscopic data respectively. For sources with more than two epochs of spectroscopic data, we notice that in most cases, the additional epochs were obtained on nearby dates to one of the epochs. As we are interested in the long term absorption line variability, we manually selected the best pair of spectroscopic epochs for each source based on the combination of maximum MJD difference between the spectroscopic observations and the highest SNR. If all the spectra from nearby epochs have high SNR, we chose the spectrum with the maximum MJD difference as the best spectrum; otherwise, we chose the spectrum with the highest SNR as the best spectrum.  For all the sources in the final sample, this selection  resulted in the earlier  epoch spectrum  to be  a part of  the SDSS-I/II  observations and the later epoch spectrum to be a part of SDSS-III observations. Hereafter, we call the earlier epoch spectrum as the SDSS-I/II spectrum and the later epoch spectrum as the SDSS-III spectrum.   We made sure that  the SDSS-I/II spectra were obtained during the time of the SDSS photometric campaign and  the SDSS-III spectra were obtained after the completion of the SDSS photometric campaign for all the sources in the final sample.  All the spectroscopic data were downloaded from the SDSS DR14 archive \footnote{http://skyserver.sdss.org/dr14/}.  The bottom panel in Fig.~\ref{fig:z_MJDdiff_plot} shows the distribution of the SN$_{1700}$ for the SDSS-I/II and SDSS-III spectra in the final sample. The dashed/blue vertical line denotes the cut, SN$_{1700}$ = 4, for selecting the high-SNR sample.

\subsection{Photometric observations}
The SDSS \citep{york00} provides  deep (r $<$ 22.5) photometry of $\sim$ 12,000 deg$^2$ in the Northern galactic cap (NGC) and $\sim$ 290 deg$^2$ in the Southern galactic cap (SGC) in five passbands \citep[ugriz;][]{fukugita96} accurate to 0.02 mag. The photometric data were obtained between July 2000 and July 2008.  The small area in the SGC (22h24m $<$ $\alpha_{J2000.0} <$ 04h08m, -1.27$^o$ < $\delta_{J2000.0} <$ +1.27$^o$), called Stripe 82 has more than 60 epochs of photometric observations available with a cadence that effectively samples time-scales ranging from days to years. { \citet{chelsea10} (hereafter, ``CM10" and the corresponding catalog presented in the CM10 paper as  the ``CM10 catalog") modeled the time variability of $\sim$9000 spectroscopically confirmed quasars in the Stripe 82 using a damped random walk (DRW) process.} CM10 used the improved photometric calibration magnitudes given in \citet{ivezic07} for their analysis.  The SDSS light curve data used in this study are mainly compiled from the database provided by CM10\footnote{http://faculty.washington.edu/ivezic/macleod/qso\_dr7\\/Southern\_format\_drw.html} (See CM10 paper for more details regarding the SDSS light curves).  For 12 sources that did not have a match in the CM10 database, we obtained the light curve information from the SDSS Data Release 14 (DR14) database. 
\begin{figure*}
    \centering
    \includegraphics[scale=0.6]{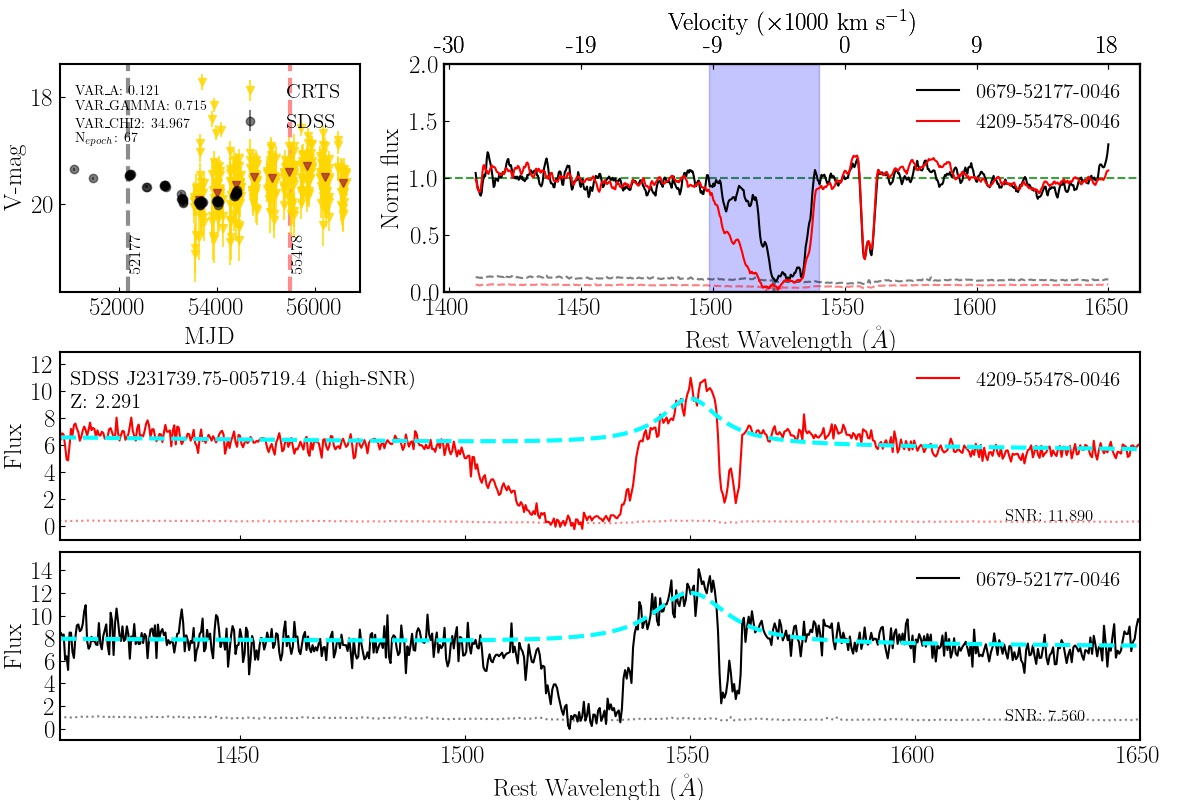}
    \caption{Example plot demonstrating the measurement of absorption trough parameters. The top left panel shows the light curves from the SDSS(black circles) and CRTS (yellow triangles) surveys. The binned CRTS light curve days is shown as red triangles. The black and red vertical lines correspond to the  MJDs corresponding to the first and second spectroscopic epochs.  The bottom and middle panels show the two epoch spectra (black/red solid line) together with the spectral uncertainties (black/red dotted line) and the best-fitting continuum (cyan/dashed line). The top right panel shows an overplot of the normalized spectra for the two epochs with the absorption troughs marked by the shaded region. }
    \label{fig:example_cont_fit}
\end{figure*}

We also queried the CRTS survey data release 2 (DR2) database \footnote{http://nesssi.cacr.caltech.edu/DataRelease/}  to obtain the  V-band light curves of quasars in our sample. CRTS obtains data in an unfiltered setup, and the resulting open magnitudes are converted to Johnson's V-band magnitude using stars present in the same field. CRTS has a typical cadence of 21 days, and in each visit, it obtains four exposures separated by 10 minutes. Except for four sources, we obtained CRTS light curves for all the sources in the final sample. CRTS observations were carried out between July 2005 to July 2013. However, the SDSS-I/II spectra were obtained much before 2005 and the SDSS-III spectra were obtained during the CRTS monitoring campaign for a majority of the sources in our sample. As the continuum variations are thought to be driving the BAL variations, it is ideal to have the photometric observations precede the later epoch of the spectroscopic observation. In this regard, CRTS data are not  ideally suited  to probe the correlation between the absorption line variability and the continuum variability for our sample. With that caveat in mind, we explore the correlations between the CRTS continuum variability parameters and the absorption line variability parameters.

A potential problem in the correlation studies involving absorption line and continuum variability is regarding the time-scales of the observations.  If the photometric and spectroscopic observations are carried out at disjoint time periods, the correlation analysis may not yield any useful results. With the earlier spectrum obtained during the photometric monitoring period and the later spectrum obtained after the photometric monitoring period, our SDSS dataset  is ideally suited to correctly probe the correlations between the absorption line variability and the continuum variability.   The top panel in Fig.~\ref{fig:z_MJDdiff_plot} shows the distribution of minimum and maximum MJDs of the photometric observations and the distribution of MJDs of SDSS-I/II and SDSS-III spectroscopic observations.  Note that  the SDSS-I/II spectroscopic MJDs  are distributed between the histograms of minimum  and maximum MJDs of photometric observations and the SDSS-III spectroscopic MJDs  are distributed after the histogram of maximum MJDs of photometric observations.

\section{Variability analysis}
We briefly describe the procedure  followed to measure the different parameters characterizing the absorption line variability and the continuum variability of sources in our sample.

\subsection{Absorption line variability parameters}

We followed the procedures listed in \citet{grier16} and \citet{gibson09b} to fit the continuum of spectra in our sample. We first excluded the pixels marked by the BRIGHTSKY  and AND\_MASK bitmask flags. Then, we corrected the spectra for Galactic extinction using the reddening curve from \citet{car89} and the $A_{\textrm{v}}$ values from \citet{sch11} assuming $R_{\textrm{v}}$~$=$~3.1. We used the visual inspection redshifts from the DR12Q catalog to convert the spectra to the rest-frame of the quasar. As our aim was to measure the \civ\ absorption line variability, local continuum was fit to the rest wavelengths between 1410 \AA\ and 1650 \AA. Together with a power-law model for the continuum, the emission lines were modeled using three different types of line profiles for each spectrum: 1) A Voigt profile, 2) A double-Gaussian profile, and 3) a 4th order Gauss-Hermite profile. Our automatic fitting algorithm makes use of an iterative fitting technique \citep[e.g.,][]{gibson09b} which masks wavelength bins that differ by more than 3$\sigma$ from the fit in the previous iteration. The best line profile (Voigt, double-Gaussian, or Gauss-Hermite) was chosen based on the reduced $\chi^2$  values ($\chi^2_{red} \sim$ 1 ).  We forced the same line profile for all the epochs of a source. A double-Gaussian profile was selected as the best line profile for 75 percent of the sources and a  Voigt line profile was chosen for the remaining sources.  To quantify the uncertainties in the continuum fitting procedure, we employ an iterative Monte-Carlo method as described in \citet{grier16}.  For each spectrum, we refit the continuum 100 times wherein each iteration, the flux of each pixel in the spectrum is altered by a random Gaussian deviate multiplied by its uncertainties. The standard deviation of the 100 iterations is taken as a measure of the uncertainty in the continuum fitting.  Finally, we normalized each spectrum by its best-fitting continuum.

{ We used the normalized spectra to identify the locations of the absorption troughs automatically.}  The trough identification algorithm is explained in detail in our forthcoming paper, Vivek et al. 2019 (in preparation) which focuses on studying BAL variability in a large sample of SDSS-IV selected quasars. Briefly, the algorithm searches for contiguous regions of \civ\ absorption with normalized flux $<$ 0.9 and absorption trough width $>$ 500 \kms that lies within a velocity range 0 $> v$  $>$ -30000 \kms\ on smoothed (Savitzy-Golay smoothing using 5 pixels) and normalized spectra.  We did not distinguish between BALs and mini-BALs as they are known to be a related phenomena. We excluded NALs from our analysis as they can also arise from the intervening absorption systems. Additionally, we  also excluded the blue-shifted absorption troughs in this analysis.     The velocity limits for each trough, $v_{max}$ and $v_{min}$ were identified as the velocities beyond which the normalized flux rises above 0.9. Our procedure identifies a single \civ\ absorption trough for 56 sources in the final sample. Among the remaining sources with multiple \civ\ absorption lines, 15 sources have two troughs, six sources have three troughs, and one source has four troughs. {  Each trough is considered independently for quasars with multiple absorption troughs.} Fig.~\ref{fig:example_cont_fit} demonstrates our continuum fitting and absorption trough identification procedure for one of the quasars in our sample. The bottom and middle panels show the two epoch spectra (black/red solid line) together with the spectral uncertainties (black/red dotted line) and the best-fitting continuum (cyan/dashed line). The top right panel shows an overplot of the normalized spectra for the two epochs with the absorption troughs marked by the shaded region. 

We further measured the standard trough parameters like equivalent (EW) width, absorption index \citep[AI;][]{hall02,trump06}, mean trough depth (d$_{avg}$) and maximum trough depth (d$_{max}$; largest average depth over a sliding 3-pixel-wide window across the trough).  { For each trough, the d$_{avg}$ and d$_{max}$ is computed over the velocities between the $v_{max}$ and $v_{min}$.  AI is defined as,
\begin{equation}
 \textrm{AI} = \int_{0}^{29,000}[1-f(v)] C^{\prime} dv     
\end{equation}
where f(v) is the normalized flux. The constant $C^{\prime}$= 1 in all regions where f(v) $<$ 0.9 is continuously for at least 1000 \kms\ and $C^{\prime}$= 0 otherwise.
We  measured the absorption index rather than the standard BAL strength metric, balnicity index \citep{weymann91} as our analysis also includes mini-BAL absorption troughs.}

Almost all previous BAL variability studies quantified the BAL variability in terms of two metrics: change in equivalent width  ($\Delta$EW) and fractional change in equivalent width ($\Delta$EW/$<$EW$>$). Following the same methodology, we used the following equations to measure the two parameters of  absorption variability and their corresponding uncertainties,
 \begin{equation}
 %\begin{center}
 \begin{aligned}
 \Delta EW &= EW_2 -EW_1 \\
 \sigma_{\Delta EW} &= \sqrt{\sigma_{EW_2}^2 + \sigma_{EW_1}^2},
 \end{aligned}
% \end{center}
 \end{equation}
 
 \begin{equation}
 \centering
 \begin{aligned}
 \frac{\Delta EW}{\langle EW \rangle} &= \frac{EW_2 -EW_1}{0.5\times(EW_2 + EW_1)} \\
 \sigma_{\frac{\Delta EW}{\langle EW \rangle}} &= \frac{4\times(EW_2 \sigma_{EW_1}+ EW_1 \sigma_{EW_2})}{(EW_2 + EW_1)^2},
 \end{aligned}
 \end{equation}
 
 Table~\ref{appendix_table1} lists the absorption line variability parameters and other absorption trough properties of sources in the final sample.  

\subsection{Continuum variability parameters}
{ Firstly, we tested the correctness of DR12Q structure function parameters, VAR\_A, and VAR\_GAMMA by computing the A and $\gamma$ values of sources in our final sample using the SDSS light curves. We see that our measurements of A and $\gamma$ are highly correlated     with the VAR\_A, VAR\_GAMMA parameters from DR12Q catalog with a Spearman's rank correlation coefficient above 0.7.  }

Apart from the DR12Q structure function parameters, VAR\_A and VAR\_GAMMA,  DR12Q catalog also provides  VAR\_CHI2,  which is the reduced $\chi^2$ when the light curve is fit with a constant. The VAR\_CHI2 parameter is high for sources with large fluctuations in the continuum flux. A potential issue with the DR12Q VAR\_A  parameter is that it is measured in the observer time-scale. 
As has been  explained in the previous sections, the photometric epochs for the final sample sources are nicely distributed between the two spectroscopic observations. 
%Nevertheless, we chose to use these parameters as they are estimated from SDSS photometric observations that are ideally suited for our analysis. In all the sources in the final sample, the first spectroscopic epoch was obtained in the beginning years of photometric campaign and the second spectroscopic epoch was obtained much after the completion of the photometric campaign. 
So, it is reasonable to assume that the DR12Q parameter, VAR\_A is a  good measure of the typical variability of the quasar continuum between the spectroscopic epochs though it is measured in the observer frame.    %These three DR12Q parameters form the primary parameters characterizing the continuum variability.

\citet{kelly09} and \citet{koz10} modeled the optical variability of quasars using a damped random walk (DRW) mechanism. In these models, the long term variability of a quasar is characterized by the SF$_\infty$ parameter which is the asymptotic value of the structure function at long time-scales. We matched the sources in our sample to the CM10 catalog which contains the DRW parameters estimated for the Stripe 82 quasars. CM10 catalog has DRW parameters listed for 66 sources in the final sample and 43 sources in the high-SNR sample.  However,  the DRW parameters are only well constrained for 45 sources in the final sample and 27 sources in the high-SNR sample (based on the likelihood of DRW solution as given in CM10). In this study, we only estimated the  SF$_\infty$ parameter for sources with well-constrained DRW parameters.
\begin{table*}
    \centering
    \begin{tabular}{p{3.5cm}p{2.3cm}p{2.3cm}p{2.3cm}p{2.3cm}}
    \hline
                 & 	\multicolumn{2}{c}{$\Delta$EW}		&	\multicolumn{2}{c}{$\Delta$EW/$\langle$EW$\rangle$} \\ %  %

Continuum parameters          & 	Final sample 	&high-SNR sample		&	Final	sample	& high-SNR sample \\
                   & 	$\rho_s$ (p-value)	& $\rho_s$ (p-value)	&	$\rho_s$ (p-value)	& $\rho_s$ (p-value)  \\
\hline
%\hline

VAR\_A         			        & \bf{0.244 (0.011)} &	\bf{0.455 (0.000)}	& \bf{0.252 (0.009)} & \bf{0.492 (0.000)}\\ 
VAR\_GAMMA                	    & 0.038 (0.695) &	-0.042 (0.738)	& 0.044 (0.651) & -0.018 (0.884)       \\
VAR\_CHI2          		    & \bf{0.301 (0.002)} &	\bf{0.305 (0.012)}	& \bf{0.289 (0.002)} & \bf{0.323 (0.008)}\\
%%VAR-MATCHED          		    & 0.202 (0.036) &	0.108 (0.386)	& 0.209 (0.030) & 0.155 (0.210)        \\
SF$_{\infty}$          		    & 0.144 (0.262) &	0.236 (0.147)	& 0.148 (0.247) & 0.276 (0.089)        \\
$\sigma$(m)$_{SDSS}$            & 0.136 (0.161) &	\bf{0.335 (0.006)}	& 0.152 (0.117) & \bf{0.359 (0.003)} \\
$\sigma$(m)$_{SDSS}$      & 0.167 (0.085) &	\bf{0.409 (0.001)}	& 0.194 (0.044) & \bf{0.442 (0.000)}        \\
$\langle\Delta$m/$\Delta$t$\rangle_{SDSS}$  & 0.178 (0.065) &	\bf{0.463 (0.000)}	& 0.188 (0.051) & \bf{0.461 (0.000)}        \\
$\sigma$(m)$_{CRTS}$            & -0.046 (0.639)&	0.010 (0.933)	& -0.020 (0.838)& 0.000 (0.998)        \\
$\langle\Delta$m$\rangle_{CRTS}$      & -0.002 (0.983) &	0.071 (0.567)	& 0.025 (0.800) & 0.067 (0.590)        \\
$\langle\Delta$m/$\Delta$t$\rangle_{CRTS}$  & -0.056 (0.567)&	0.054 (0.662)	& -0.032 (0.739)& 0.054 (0.667)        \\

\hline
    \end{tabular}
    \caption{Spearman's Rank Correlation analysis of the absorption line variability and the continuum variability  parameters for the final and high-SNR sample. The statistically significant correlations are marked in bold face. }
    \label{tab:summary_SR_correlation}
\end{table*}

Besides the above continuum variability parameters compiled from different catalogs, we also measured three variability parameters prescribed by \citet{giveon99}  using the r-band SDSS and the V-band CRTS light curves. The three parameters are (1) $\sigma$(m), the standard deviation of the magnitudes (2) $\langle\Delta$m$\rangle$, median of the magnitude differences between all epochs and (3) $\langle\Delta$m/$\Delta$t$\rangle$, the median of the magnitude change in unit rest-frame time.  We noticed a large scatter in the CRTS light curves on short time-scales possibly due to calibration errors arising from unfiltered observations. So, we  median averaged each light curve in 200 days bins and used the binned light curve to compute the CRTS  variability parameters.  The three continuum variability parameters computed from SDSS  light curves are collectively termed as the "SDSS light curve variability parameters". Likewise, the "CRTS light curve variability parameters" corresponds to the group of three continuum variability parameters computed using the CRTS light curves. Table~\ref{appendix_table2} lists the continuum variability parameters of all  sources in the final sample.  

Continuum variability parameters, VAR\_A, VAR\_CHI2, $\sigma$(m), $\langle\Delta$m$\rangle$ and $\langle\Delta$m/$\Delta$t$\rangle$ are measures of the amplitude of the variability,   whereas VAR\_GAMMA is a measure of the dependence of variability on the time-scale. The parameter SF$_\infty$ has a dependence on both the amplitude and the time-scale of the variability. As the different continuum variability parameters are compiled/measured from different catalogs/ observations, we first explored the correlation of these parameters between themselves. The  Spearman's rank correlation coefficients and the associated p-values between the different continuum variability parameters are listed in Table~\ref{tab:appendix_correlationtable}. The variability amplitude parameters VAR\_A, VAR\_CH12, SF$_{\infty}$  and the  SDSS light curve variability parameters are all derived from the same SDSS photometric observations and should be highly correlated with each other.  As expected, DR12Q parameter VAR\_A is highly correlated ($\rho_s$ > 0.6) with  VAR\_CHI2, SF$_{\infty}$ and the three SDSS light curve variability parameters.  We find no significant correlation ($\rho_s$ < 0.3) between the CRTS light curve variability parameters and the other continuum variability parameters even though the three CRTS  light curve variability parameters are highly correlated between themselves. The lack of correlation between the SDSS light curve variability parameters and the CRTS light curve variability parameters may be attributed to the difference in the time-scales of the SDSS and CRTS photometric observations. { To further explore the lack of significant correlation between the SDSS and CRTS light curve variability parameters, 	we compiled the r-band light curves from the Palomer Transient Factory (PTF) Data Release 3 (DR3)  for 56 sources in the final sample. 
The SDSS  light curves were obtained between MJDs 51500 and 54500, where as the CRTS light curves were obtained between MJDs 53500 and 56500. The PTF light curves in turn were obtained between MJDs 55000 and 56800. Clearly, the PTF light curves have more overlap with the CRTS light curves than the SDSS light curves. The rms value of the PTF r-band magnitudes (obtained from the PTF DR3 catalog) is adopted as the measure of PTF light curve variability.  We computed the correlation strengths between the SDSS, CRTS and PTF light curve variability parameters and find that the PTF light curve variability parameter is highly correlated ($\rho_s \sim$ 0.5) with  the CRTS light curve variability parameters as compared to the SDSS light curve variability parameters ($\rho_s \sim$ 0.3).  This is consistent with our earlier measurement that CRTS and SDSS light curve variability parameters are weakly correlated ($\rho_s \sim$ 0.3). The lack of correlation between the SDSS and CRTS continuum variability parameters suggests that the timescales over which the continuum variability are estimated also plays an important role in the correlations between continuum variability and absorption line variability. In this regard, CRTS continuum variability parameters may not be best suited to explore the correlation between absorption line variability and continuum variability for our sample. }

\begin{figure*}
    \centering
    \includegraphics[scale=0.6]{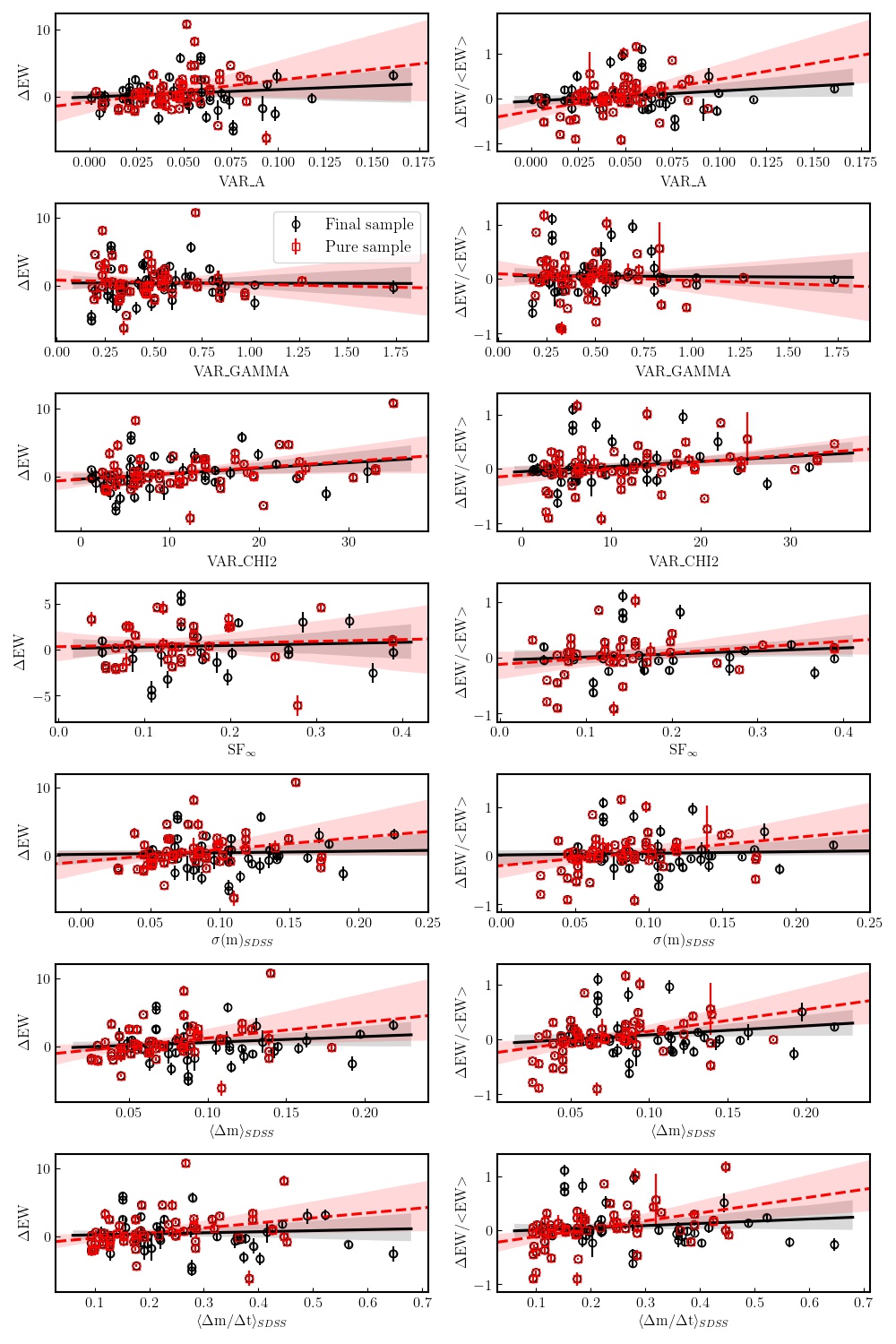}
    \caption{Left Panels : Scatter plots showing the variation of $\Delta$EW with the continuum variability parameters.  Right Panels: Scatter plots showing the variation of $\Delta$EW/$\langle$EW$\rangle$ with the continuum variability parameters. In all the panels, the black circles and the red squares correspond to final and high-SNR sample sources. Results of linear regression for the final and high-SNR sample are shown as solid/black and dashed/red lines. The shading corresponds to 95 percent confidence intervals of the fit. }
    \label{fig:correlation_plot}
\end{figure*}

\section{Statistical analysis}
{ Previous studies have reported that  BAL variability is positively correlated with the rest-frame time-scale and the absorber velocity,  and inversely correlated with the trough depth \citep{gibson09,capellupo11,filiz13,vivek14,welling14,vivek16}. %In this work, we also confirm the previously found  BAL variability trends with rest-frame time-scale, absorber velocity and trough depth for our sample of sources. 
In the rest of this paper, we mainly focus on investigating the correlation between the absorption line variability and the continuum variability.}

The left and right panels of Fig.~\ref{fig:correlation_plot} show  the variation of $\Delta$EW and $\Delta$EW/$\langle$EW$\rangle$ as a function of the continuum variability parameters VAR\_A, VAR\_CHI2, SF$_{\infty}$, $\sigma$(m)$_{SDSS}$, $\langle\Delta$m$\rangle_{SDSS}$ and $\langle\Delta$m/$\Delta$t$\rangle_{SDSS}$. The final and high-SNR sample sources are shown as black circles and red squares respectively. The best-fitting linear regression line with 95 percent confidence intervals are shown as the dashed line and the shaded region. Clearly, the slopes of the linear regression of VAR\_A, $\sigma$(m)$_{SDSS}$, $\langle\Delta$m$\rangle_{SDSS}$ and $\langle\Delta$m/$\Delta$t$\rangle_{SDSS}$ parameters increases for the high-SNR sample as compared to the final sample.

Table.~\ref{tab:summary_SR_correlation} lists the results of the Spearman's rank correlation analysis between the absorption line and continuum variability parameters for the final and high-SNR sample sources.
It is encouraging to note that the different correlation strengths for the absorption line variability parameters, $\Delta$EW and $\Delta$EW/$\langle$EW$\rangle$,  show similar trends.
For the final sample, we see a weak correlation ($\rho_s \sim$ 0.3) between the DR12Q continuum variability parameters VAR\_A and VAR\_CHI2 and both the absorption line variability parameters.  The correlation strength between VAR\_A and both the absorption line variability parameters improve and become moderate ($\rho_s \sim$ 0.5) for the high-SNR sample sources. Although the correlation strength for VAR\_CHI2 parameter also improves for the high-SNR sample, it does not change significantly between the high-SNR and final sample.  Additionally, we find a moderate correlation between the absorption line variability parameters and the three SDSS light curve variability parameters for the high-SNR sample. We also confirmed that the  VAR\_A parameter estimated in the rest-frame time-scales (by dividing the observer-frame VAR\_A of a source with (1+z)$^\gamma$, where z and $\gamma$ are the redshift and the VAR\_GAMMA parameter of the source) also shows a similar correlation with $\Delta$EW ($\rho_s$ = 0.442) and $\Delta$EW/$\langle$EW$\rangle$ ($\rho_s$ = 0.434).

\begin{table*}
    \centering
    \begin{tabular}{p{3.5cm}p{2.3cm}p{2.3cm}p{2.3cm}p{2.3cm}}
    \hline
                 & 	\multicolumn{2}{c}{$\Delta$EW}		&	\multicolumn{2}{c}{$\Delta$EW/$\langle$EW$\rangle$} \\ %  %

Continuum parameters          & 	Shallow trough 	&Deep	trough	&	Shallow trough	& Deep trough \\
                   & 	$\rho_s$ (p-value)	& $\rho_s$ (p-value)	&	$\rho_s$ (p-value)	& $\rho_s$ (p-value)  \\
\hline
%\hline

VAR\_A         			        & \bf{0.666 (0.000)} &	{0.242 (0.169)}	& \bf{0.666 (0.000)} & {0.237 (0.176)}\\ 
VAR\_GAMMA               	    & -0.004 (0.984) &	-0.062 (0.729)	& -0.016 (0.932) & -0.080 (0.650)       \\
VAR\_CHI2        		    & \bf{0.398 (0.022)} &	{0.243 (0.166)}	& \bf{0.371 (0.034)} & {0.243 (0.165)}\\
%%VAR-MATCHED          		    & 0.202 (0.036) &	0.108 (0.386)	& 0.209 (0.030) & 0.155 (0.210)        \\
SF$_{\infty}$          		    & 0.433 (0.072) &	0.020 (0.931)	& 0.336 (0.173) & 0.078 (0.737)        \\
$\sigma$(m)$_{SDSS}$            & \bf{0.446 (0.009)} &	{0.241 (0.170)}	& \bf{0.430 (0.012)} & {0.200 (0.256)} \\
$\langle\Delta$m$\rangle_{SDSS}$      & \bf{0.533 (0.001)} &	{0.316 (0.069)}	& \bf{0.529 (0.002)} & {0.292 (0.094)}        \\
$\langle\Delta$m/$\Delta$t$\rangle_{SDSS}$  & \bf{0.643 (0.000)} &	{0.286 (0.101)}	& \bf{0.613 (0.000)} & {0.217 (0.219)}        \\
$\sigma$(m)$_{CRTS}$            & -0.017 (0.926)&	0.049 (0.782)	& 0.040 (0.824)& -0.102 (0.564)        \\
$\langle\Delta$m$\rangle_{CRTS}$      & 0.044 (0.810) &	0.129 (0.464)	& 0.081 (0.654) & 0.005 (0.979)        \\
$\langle\Delta$m/$\Delta$t$\rangle_{CRTS}$  & 0.017 (0.923)&	0.096 (0.589)	& -0.048 (0.793)& 0.003 (0.986)        \\

\hline
    \end{tabular}
    \caption{Spearman's Rank Correlation analysis of the absorption line variability and the continuum variability  parameters for the shallow and deep trough sub-samples of the 45 sources in the high-SNR sample. The statistically significant correlations are marked in bold face.}
    \label{tab:summary_SR_correlation_SD}
\end{table*}

The VAR\_GAMMA parameter, a measure of the time-scale dependence of continuum variability has no significant correlations with the absorption line variability parameters for both the final and high-SNR sample. 
SF$_{\infty}$ parameter also does not show any correlations with both the absorption line variability parameters. The lack of correlation of SF$_{\infty}$ may be explained by the dependence of  SF$_{\infty}$ on the time-scale of the variability. SF$_{\infty}$ parameter is correlated with both VAR\_A and VAR\_GAMMA with a  correlation strength of 0.72 and 0.44 respectively (See, Table~\ref{tab:appendix_correlationtable}).  As we do not find any correlation between the VAR\_GAMMA parameter and the absorption line variability parameters, we suspect that the correlation strengths between SF$_{\infty}$ and the absorption variability parameters are diluted by the dependence of SF$_{\infty}$ on VAR\_GAMMA.
As discussed before, the CRTS light curve variability parameters are not the best parameters for probing the connection between absorption line and continuum variability for our sample. We do not see any correlation between the three CRTS light curve variability parameters and the absorption line variability parameters.

In the correlation analysis, we see that the correlation strengths of statistically significant (p-value < 0.05) continuum variability parameters always improve for the high-SNR sample as compared to the final sample.   The average SNR for the final and the high-SNR sample spectra are 8 and 12 respectively.   It is clear that the low spectral SNR of many sources in the final sample spectra is responsible for the weak/no correlation of the VAR\_A, VAR\_CHI2 and the SDSS light curve variability parameters with the absorption line variability parameters.

{ We used Monte Carlo (MC)  methods to test the effect of systematic errors on the correlation between the three SDSS, CRTS continuum variability parameters, and the absorption line variability parameters. We used the bootstrap error as an estimate of the uncertainty in the continuum variability parameters. In each iteration of the MC, we altered the values of the parameters by a random Gaussian deviate multiplied by its uncertainties and computed the correlation coefficients and the associated p-values.  The median of 100 iterations is adopted as the final correlation coefficient and the p-value.     This analysis yields comparable correlation strengths for the previously found statistically significant correlations, i.e., between SDSS light curve variability parameters and the absorption line variability parameters. The standard deviation of the statistically significant correlation coefficients is found to be below 0.06. None of the CRTS light curve variability parameters show significant correlations with the absorption line variability parameters.}

\subsection{Effect of saturation?}
Saturation in the absorption lines can weaken the correlation between the absorption line variability and continuum variability as a saturated absorber might not respond to the continuum fluctuations.  
Several previous studies have reported that BAL  profiles suffer from saturation \citep[see,][]{arav97,hamann98,arav01,leighly09,borguet12,xu18}.  The depth of a saturated absorption line is mostly determined by the covering fraction of the background ionizing source by the absorbing material. The absorption lines arising out of a saturated absorber can be non-black if the absorbing gas does not completely cover the background source. \citet{filiz13} showed that the equivalent width of deeper absorption troughs varies less compared to the shallower absorption troughs.

To probe the effect of saturation in our correlation studies, we split the \civ\ absorption troughs in the high-SNR sample sources into two sub-samples according to the maximum trough depth parameter, d$_{max}$. As the sources in our sample do not exhibit any dramatic trough variabilities like BAL emergence/ disappearance, we used the average value of the two epochs, $\langle$d$_{max}\rangle$ to split the sample. The minimum and maximum values of the  $\langle$d$_{max}\rangle$ for the high-SNR sample sources are 0.3 and 1 respectively. So, we divided the sample into a shallow trough sub-sample and a deep trough sub-sample depending on whether the $\langle$d$_{max}\rangle$ is below or above 0.7. The shallow trough sub-sample has 33 \civ\ absorption troughs, and the deep trough sub-sample has 34 \civ\  absorption troughs.

Table.~\ref{tab:summary_SR_correlation_SD} shows the results of the correlation analysis between the absorption line variability and continuum variability parameters for the shallow and deep trough sub-sample. Interestingly, the correlation strengths of all the statistically significant continuum variability parameters (i.e., VAR\_A, VAR\_CHI2, $\sigma$(m)$_{SDSS}$, $\langle \Delta$m$\rangle_{SDSS}$  and $\langle\Delta$m/$\Delta$t$\rangle_{SDSS}$ ) improve substantially  ($\rho_s \sim$ 0.6) for  the shallow trough sub-sample, whereas the deep trough sub-sample does not show any correlation between the absorption line variability and continuum variability parameters. It is clear that the saturation of the absorption troughs has a pronounced effect in determining the correlation strength between the absorption line variability and the continuum variability parameters. 

\begin{figure*}
    \centering
    \includegraphics[scale=0.6]{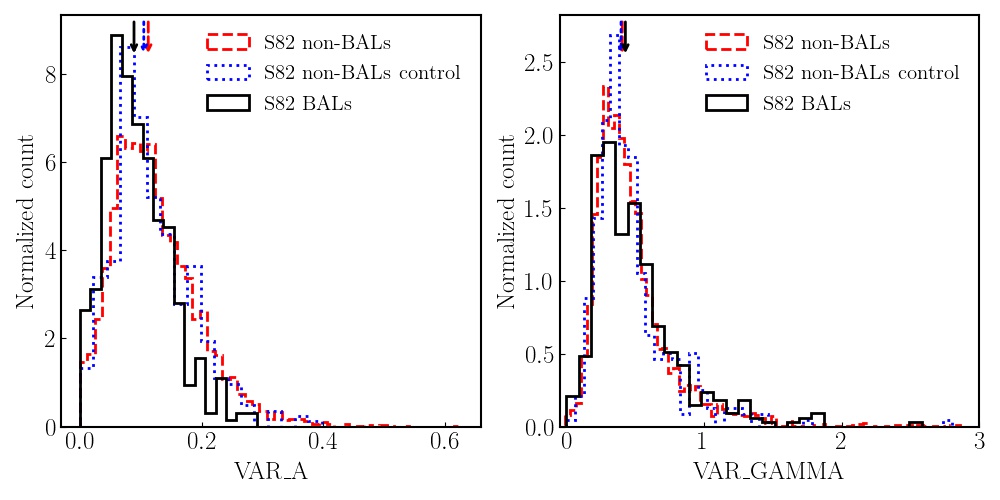}
    \caption{Normalized histograms of VAR-A (left panel) and VAR-GAMMA (right panel) parameters for the Stripe 82 BAL quasar (solid/black line), Stripe 82 non-BAL quasar (dashed/red line) and Stripe 82 non-BAL quasar control (dotted/blue line) samples. The vertical arrows mark the median values for the three samples. }
    \label{fig:BALs_vs_nonBALs}
\end{figure*}

\section{Discussion and Conclusion}

Major evidence for ionization driven absorption line variability for a large sample of quasars are presented in \citet{filiz13,wang15,he17,lu17,lu18}. Using a sample of 291  quasars, \citet{filiz13} used the coordinated variations between different absorption components as a diagnostic for ionization driven BAL variability. \citet{wang15} used the emission lines and the flux at 1400 \AA\ as a measure of continuum flux. They defined a new measure of the coordinated absorption line and continuum variability, the concordance index ( +1 represents the cases where both the absorption line and  the continuum become stronger or weaker together; -1 represents the cases where the absorption line becomes stronger when the continuum weakens or vice versa; 0 otherwise) and found that a large fraction of the sources in their sample have a  concordance index of -1. { \citet{wang15} support the idea that absorption-line variability is driven mainly by changes in the gas ionization in response to continuum variations.} In a different approach, \citet{he17} defined the ratio of equivalent widths of \siiv\ and \civ\ as an indicator of the ionization state of the absorber. They concluded that BAL variabilities in 80 percent cases are driven by changes in the continuum.  \citet{lu17} adopted the method of \citet{wang15} to study photo-ionization driven absorption line variability in a sample of 40  narrow absorption line quasars and reported a strong correlation between absorption line and continuum variability.  Using the spectral flux at 1450 \AA\ and 1700 \AA, \citet{lu18} reported moderate correlations between the fractional changes in the equivalent width and the continuum for a large sample of  \civ\ and \siiv\ BALs. \citet{stern17} presented an extreme case of BAL variability in one quasar where several absorption components disappeared while the quasar continuum brightened dramatically. 

In contrast to the above results, \citet{wildy14} found no correlation between the BAL variability and the luminosity of the quasar and suggested that ionizing continuum changes do not play a significant role in BAL variability.  Using CRTS photometric observations, \citet{vivek14,welling14,vivek16} did not find any significant trend between the variations of absorption lines and continuum flux.   

In all the previous studies involving continuum variability, estimates of continuum were measured either from the spectra itself or from the CRTS observations. As found in this study, estimates using CRTS light curves may not be the best metric of continuum variability that can probe the connection between the continuum and the absorption lines for a sample of quasars that have the earlier spectroscopic data from SDSS-I/II. In this work, we directly measured the continuum variability from the SDSS photometric observations that were obtained between the two spectroscopic observations used for measuring the absorption line variability.

 The continuum variability parameters used in this study are not sensitive to the direction of continuum variation, i.e., whether the continuum brightens or weakens. It is well known that  absorption line response to ionization changes is not monotonic \citep[see Fig.~2 of][]{hamann97}. Depending on the value of the ionization parameter, the column density of the ions can increase or decrease for an increase in the ionizing radiation. Our choice of continuum variability parameters ensures that the    correlation between the absorption line and the continuum variability is not affected by the complex relation between ionization parameter and column density.

\citet{wang15} reported that the equivalent width of the BALs decrease (increase) statistically when the continuum brightens (dims).  In our analysis, we notice that the photometric magnitudes of several sources do not vary monotonically with time (See the top left panel of Fig.~\ref{fig:example_cont_fit} for an example). The light curves typically exhibit turnovers in magnitude variations consistent with the DRW model of quasar flux variability \citep{kelly09,chelsea10}. Hence, we could not probe the trend of BAL equivalent width decrease (increase)  when the continuum brightens (dims). As mentioned before, our continuum variability parameters rather point to the overall nature of the quasar to be variable in its continuum flux. 

In our correlation analysis, we see only a moderate correlation between the absorption line and continuum variability even for the high-SNR sample. We have shown that saturation in the absorption troughs  weakens the correlation between the continuum and the absorption line. Two other important effects should be considered before interpreting the above correlation strengths. Firstly, the r-band and V-band light curves used in this study may not be a representative of the actual ionizing photons.  It is well known that the short wavelength photons of AGN vary faster than the long wavelength photons \citep [for eg.,][]{nandra98}.  It is possible that changes in the ionizing UV continuum are much more than the observed changes in the V-band \citep{welsh11} and thus,  may be more strongly correlated with the absorption line variability. However, the long and short wavelength light curves are known to be correlated with  a time-lag, $\tau \propto \lambda^{\frac{4}{3}}$ \citep[for e.g.,][]{cackett07}. The indirect dependence of the absorption line on the V-band continuum may contribute to weaken their correlation.
{ Secondly, the ionization state of the absorbing gas only retains the memory of the ionizing continuum for about a recombination time scale. The ionization changes should happen on time-scales larger than the recombination time-scale for driving the absorption line variations. Small time-scale continuum variations will smear the correlation between the continuum and absorption line changes \citep{wang15}.} Considering the above two aspects together with the saturation effects, it is highly probable that the actual correlation between the absorption line and continuum changes is even stronger.  

%Our study is not best suited to probe the short time-scale absorption line variability as we preferentially select the two spectroscopic epochs with maximum MJD separation.  
%Using the short time-scale ($\sim$ few days) BAL variability studies, \citet{capellupo12,grier15} have constrained the electronic density of the absorber to be $>$ 10$^4$  cm$^{-3}$. The average rest-frame time-scale between the spectroscopic observations for our sample is $\sim$ 800 days. The variability time-scale and the recombination rate \citep[computed using the coefficients in][for a 10$^5$ K temperature cloud]{badnell06} yield a lower limit on the electron density n$_e$ $>$ 5$\times$10$^3$ cm$^3$ consistent with previous studies.

We further explored the question if the continuum variations in   BAL quasars are any different from that of non-BAL quasars. We compared the DR12Q continuum variability parameters, VAR\_A, and VAR\_GAMMA of a sample of BAL quasars (sample size: 374) and non-BAL quasars (sample size: 3638) in the Stripe 82 region. We only include quasars that have at least ten photometric epochs in their light curves. As the sample sizes are considerably different, we also constructed a control sample of non-BAL quasars. For every quasar in the Stripe 82 BAL quasar sample,  we identified a non-BAL quasar from the Stripe 82 non-BAL quasar sample using the k-nearest neighbor algorithm, which minimizes the distance in the redshift, absolute luminosity plane.   Fig.~\ref{fig:BALs_vs_nonBALs} shows the normalized histograms of VAR\_A (left panel) and VAR\_GAMMA (right panel) parameters for the Stripe 82 BAL quasar (solid/black line), Stripe 82 non-BAL quasar(dashed/red line) and Stripe 82 non-BAL quasar control (dotted/blue line) samples. The vertical arrows mark the median values for the three samples. We do not see any marked difference in the distribution of VAR\_A and VAR\_GAMMA between the BAL quasar sample and the two non-BAL quasar samples. We also confirmed this trend using the full sample of quasars in the DR12Q catalog.  We conclude that the continuum variability in  BAL quasars is not different from that of non-BAL quasars. As mentioned in the introduction section, continuum variations in a BAL quasar can be caused due to the intrinsic variations of the quasar flux or due to changes in the shielding gas. As non-BAL quasars do not require the presence of shielding gas between the central source and the BAL absorber, the similarity in the distribution of continuum variability parameters between BAL quasars and non-BAL quasars suggests that the continuum variations driven by the shielding gas changes are not a dominant mechanism. However, the shielding gas may only modify the far-UV BAL ionizing photons without altering the visible-band flux \citep[for eg., see Fig.8 of][]{vivek18}. In such scenarios, we cannot disentangle the effects of the two modes of continuum variability by comparing the continuum variability parameters of a BAL  and non-BAL quasar sample.

In this work, we probed the connection between the \civ\ absorption line variability and continuum variability using a sample of 78 BAL quasars in the Stripe 82 region that have well sampled photometric as well as spectroscopic observations. The main conclusions from our analysis are the following,
\begin{enumerate}
    \item We find evidence for weak correlations between the absorption line variability parameters, $\Delta$EW and $\Delta$EW/$\langle$EW$\rangle$, and the continuum variability parameters, VAR\_A, VAR\_CHI2, $\sigma$(m)$_{SDSS}$,$\langle \Delta$m$\rangle_{SDSS}$ and  $\langle\Delta$m/$\Delta$t$\rangle_{SDSS}$ for the final sample sources.
    \item The statistically significant correlation strengths of the continuum variability parameters  become moderate ($\rho_s \sim$ 0.5) for the high-SNR sample as compared to the final sample. 
    \item Using two sub-sets of the high-SNR sample differing on the absorption trough depth, we find that saturation effects play an important role in the correlation between absorption changes and continuum changes. While the correlation strengths of the shallow trough sub-sample improve substancially ($\rho_s \sim$ 0.6), the deep trough sample does not show any correlations between the absorption line variability parameters and the continuum variability parameters.
    \item Considering other effects that can also smear the correlation, we conclude that the actual correlation between the absorption line and continuum variability is even stronger.
    \item We do not find any statistical difference in the continuum variability properties of BAL quasars and non-BAL quasars.
\end{enumerate}

SDSS-Reverberation Mapping program obtains regular repeated photometric as well as spectroscopic observations of 850 quasars with the aim of measuring their black hole masses. The sample also includes $\sim$ 90 BAL quasars for which multi-epoch photometry and spectroscopy are available over a time-scale of 5 years. Correlation studies using this high SNR data set will be valuable to settle the question of photoionization driven absorption line variability. Future time-domain astronomical surveys like SDSS-V and Large Synoptic Survey Telescope (LSST) will result in an order of magnitude increase in the sample size of such studies.

\section*{Acknowledgements}
{ We thank the anonymous referee for useful comments that helped to improve the manuscript.}
Most of the work was done when MV was visiting IUCAA, Pune. MV thanks IUCAA, Pune for its hospitality. 
MV acknowledges the help from Chelsea L. MacLeod regarding the discussion on  SDSS light curves. MV also acknowledges the help from Susmitha R. Antony for reading the manuscript.
%MV acknowledges support from NSF grant AST-1516784.

Funding for the Sloan Digital Sky Survey IV has been provided by the Alfred P. Sloan Foundation, the U.S. Department of Energy Office of Science, and the Participating Institutions. SDSS-IV acknowledges
support and resources from the Center for High-Performance Computing at
the University of Utah. The SDSS web site is www.sdss.org.

Funding for the SDSS and SDSS-II has been provided by the Alfred P. Sloan Foundation, the Participating Institutions, the National Science Foundation, the U.S. Department of Energy, the National Aeronautics and Space Administration, the Japanese Monbukagakusho, the Max Planck Society, and the Higher Education Funding Council for England. The SDSS Web Site is http://www.sdss.org/.

The SDSS is managed by the Astrophysical Research Consortium for the Participating Institutions. The Participating Institutions are the American Museum of Natural History, Astrophysical Institute Potsdam, University of Basel, University of Cambridge, Case Western Reserve University, University of Chicago, Drexel University, Fermilab, the Institute for Advanced Study, the Japan Participation Group, Johns Hopkins University, the Joint Institute for Nuclear Astrophysics, the Kavli Institute for Particle Astrophysics and Cosmology, the Korean Scientist Group, the Chinese Academy of Sciences (LAMOST), Los Alamos National Laboratory, the Max-Planck-Institute for Astronomy (MPIA), the Max-Planck-Institute for Astrophysics (MPA), New Mexico State University, Ohio State University, University of Pittsburgh, University of Portsmouth, Princeton University, the United States Naval Observatory, and the University of Washington.

Funding for SDSS-III has been provided by the Alfred P. Sloan Foundation, the Participating Institutions, the National Science Foundation, and the U.S. Department of Energy Office of Science. The SDSS-III web site is http://www.sdss3.org/.

SDSS-III is managed by the Astrophysical Research Consortium for the Participating Institutions of the SDSS-III Collaboration including the University of Arizona, the Brazilian Participation Group, Brookhaven National Laboratory, Carnegie Mellon University, University of Florida, the French Participation Group, the German Participation Group, Harvard University, the Instituto de Astrofisica de Canarias, the Michigan State/Notre Dame/JINA Participation Group, Johns Hopkins University, Lawrence Berkeley National Laboratory, Max Planck Institute for Astrophysics, Max Planck Institute for Extraterrestrial Physics, New Mexico State University, New York University, Ohio State University, Pennsylvania State University, University of Portsmouth, Princeton University, the Spanish Participation Group, University of Tokyo, University of Utah, Vanderbilt University, University of Virginia, University of Washington, and Yale University.

%%%%%%%%%%%%%%%%%%%%%%%%%%%%%%%%%%%%%%%%%%%%%%%%%%



\bibliographystyle{mnras}
\bibliography{strpaper} % if your bibtex file is called example.bib

%\begin{thebibliography}{99}

%\end{thebibliography}

%%%%%%%%%%%%%%%%%%%%%%%%%%%%%%%%%%%%%%%%%%%%%%%%%%

%%%%%%%%%%%%%%%%% APPENDICES %%%%%%%%%%%%%%%%%%%%%

\appendix
\section{variability measurements}
\begin{table*} 
         \centering 
         \fontsize{6}{9}\selectfont
         \begin{tabular}{lccccccccc}%{p{3.175cm}p{0.2cm}p{1.9cm}p{1.9cm}p{0.5cm}p{0.3cm}ccc} 
         \hline 
         SDSS name & z$^a$ & P-M-F$_1^b$ & P-M-F$_2^c$ &$\Delta$t$^{d}$ & n$_{tr}^e$ & $v_{max}$-$v_{min}^f$  & $\Delta$EW$^g$ &         $\Delta$EW/$\langle$EW$\rangle^h$ & $\langle d_{max}\rangle^i$  \\ 
        (J2000) & & & & (days)& &( km s$^{-1}$) &(\AA) &  &  \\
          \hline 
               
SDSS J210946.54$-$003633.0 &	      2.30&	1112-53180-0305&	4191-55444-0406&	    686.06&	         1&	    3134.5 -    1612.3&	      1.82$\pm$0.62&	      0.51$\pm$0.17&	      0.74$\pm$0.16 \\ 
SDSS J211200.99$-$003443.5 &	      2.50&	1112-53180-0221&	4191-55444-0262&	    647.41&	         1&	   17315.2 -   10589.6&	      0.49$\pm$0.92&	      0.06$\pm$0.11&	      0.50$\pm$0.14 \\ 
SDSS J211415.82$-$011316.5 &	      2.30&	1523-52937-0127&	4191-55444-0134&	    760.16&	         1&	    6489.2 -    3383.8&	     -0.39$\pm$0.65&	     -0.04$\pm$0.07&	      0.82$\pm$0.14 \\ 
SDSS J211518.49$+$001115.2 &	      2.09&	1112-53180-0515&	4192-55469-0570&	    740.78&	         1&	    2978.2 -    736.4&	     -0.63$\pm$1.10&	     -0.07$\pm$0.12&	      1.03$\pm$0.20 \\ 
SDSS J211829.90$+$002830.3 &	      2.62&	1112-53180-0599&	4192-55469-0724&	    633.20&	         1&	   22962.6 -   15694.4&	      0.89$\pm$1.06&	      0.13$\pm$0.15&	      0.42$\pm$0.16 \\ 
SDSS J211853.67$+$002843.9$^{\dagger}$&	      1.66&	0987-52523-0380&	5142-55825-0419&	   1241.68&	         1&	   12775.3 -    4766.5&	      3.35$\pm$0.78&	      0.31$\pm$0.07&	      0.56$\pm$0.11 \\ 
SDSS J211908.43$+$003246.3$^{\dagger}$&	      2.33&	0986-52443-0603&	4192-55469-0766&	    910.08&	         1&	    4169.8 -     825.1&	      4.69$\pm$0.14&	      0.86$\pm$0.03&	      0.61$\pm$0.03 \\ 
SDSS J212026.14$+$000735.0 &	      2.47&	1112-53180-0637&	4192-55469-0858&	    660.22&	         1&	    6904.3 -    1557.9&	      0.91$\pm$1.38&	      0.06$\pm$0.09&	      0.99$\pm$0.23 \\ 
SDSS J212204.46$-$001012.5 &	      3.47&	0987-52523-0127&	4193-55476-0424&	    660.63&	         1&	    4716.6 -    3365.9&	     -0.31$\pm$0.42&	     -0.11$\pm$0.15&	      0.58$\pm$0.15 \\ 
SDSS J212319.66$-$002245.8$^{\dagger}$&	      2.61&	0987-52523-0092&	4192-55469-0032&	    816.07&	         1&	   10612.4 -      42.5&	      1.17$\pm$1.53&	      0.05$\pm$0.06&	      1.03$\pm$0.47 \\ 
SDSS J212419.26$+$004313.4$^{\dagger}$&	      1.58&	1522-52932-0582&	5143-55828-0460&	   1124.19&	         1&	   11097.3 -    4162.0&	      2.55$\pm$0.61&	      0.09$\pm$0.02&	      1.00$\pm$0.09 \\ 
SDSS J212811.62$-$010704.8$^{\dagger}$&	      3.28&	0988-52520-0269&	4193-55476-0052&	    690.65&	         2&	   19642.0 -   17510.4&	     -0.02$\pm$0.37&	     -0.01$\pm$0.12&	      0.52$\pm$0.11 \\ 
               $ $         &	          &	 &	 &	          &	          &	   15927.6 -    9174.8&	      8.19$\pm$0.73&	      1.17$\pm$0.10&	      0.37$\pm$0.10 \\ 
SDSS J212952.14$+$003244.5$^{\dagger}$&	      1.87&	0988-52520-0469&	4194-55450-0638&	   1021.26&	         3&	   21831.2 -   18876.7&	     -0.31$\pm$0.30&	     -0.07$\pm$0.07&	      0.47$\pm$0.07 \\ 
               $ $         &	          &	 &	 &	          &	          &	   17363.5 -   15230.0&	     -1.81$\pm$0.25&	     -0.91$\pm$0.12&	      0.29$\pm$0.07 \\ 
               $ $         &	          &	 &	 &	          &	          &	    7510.2 -    3784.1&	      0.39$\pm$0.27&	      0.03$\pm$0.02&	      0.92$\pm$0.06 \\ 
SDSS J213008.34$+$002100.0$^{\dagger}$&	      2.57&	1521-52945-0518&	4193-55476-0973&	    708.77&	         2&	   16410.9 -    9971.5&	      3.43$\pm$0.61&	      0.29$\pm$0.05&	      0.61$\pm$0.10 \\ 
               $ $         &	          &	 &	 &	          &	          &	    7717.7 -     548.6&	      2.47$\pm$0.61&	      0.10$\pm$0.02&	      0.96$\pm$0.10 \\ 
SDSS J213138.07$-$002537.8$^{\dagger}$&	      1.84&	0989-52468-0273&	4194-55450-0266&	   1050.74&	         1&	    2919.1 -    1418.3&	      0.55$\pm$0.37&	      0.17$\pm$0.11&	      0.74$\pm$0.12 \\ 
SDSS J213648.17$-$001546.6$^{\dagger}$&	      2.17&	0989-52468-0104&	4195-55452-0348&	    939.84&	         1&	    5652.0 -    2882.2&	     -4.27$\pm$0.17&	     -0.54$\pm$0.02&	      0.73$\pm$0.04 \\ 
SDSS J213901.65$-$004925.2 &	      3.66&	0989-52468-0011&	4195-55452-0256&	    640.34&	         1&	   18983.5 -   11548.0&	     -3.24$\pm$0.93&	     -0.24$\pm$0.07&	      0.64$\pm$0.17 \\ 
SDSS J214107.19$+$001213.4 &	      3.65&	0989-52468-0635&	4195-55452-0874&	    642.27&	         2&	   11474.4 -    8095.6&	     -0.28$\pm$0.60&	     -0.04$\pm$0.09&	      0.59$\pm$0.11 \\ 
               $ $         &	          &	 &	 &	          &	          &	    2820.7 -     952.7&	      0.97$\pm$0.35&	      0.20$\pm$0.07&	      0.76$\pm$0.11 \\ 
SDSS J214113.05$-$003545.9$^{\dagger}$&	      2.23&	0990-52465-0188&	4196-55478-0466&	    933.68&	         1&	    7695.3 -    4038.2&	     -1.93$\pm$0.23&	     -0.30$\pm$0.04&	      0.59$\pm$0.05 \\ 
SDSS J214227.48$+$005651.7$^{\dagger}$&	      3.63&	0990-52465-0412&	4196-55478-0594&	    650.76&	         1&	    5084.4 -    1418.2&	     -0.78$\pm$0.39&	     -0.09$\pm$0.05&	      0.73$\pm$0.08 \\ 
SDSS J214456.61$+$005926.2 &	      1.70&	1108-53227-0539&	5146-55831-0466&	    964.44&	         1&	    4911.2 -    1184.0&	      3.17$\pm$0.80&	      0.23$\pm$0.06&	      1.02$\pm$0.17 \\ 
SDSS J214701.79$+$001643.3 &	      1.89&	1107-52968-0354&	5146-55831-0382&	    991.11&	         1&	   16725.0 -   13420.0&	      2.97$\pm$0.48&	      0.82$\pm$0.13&	      0.36$\pm$0.11 \\ 
SDSS J214821.43$-$003705.8 &	      2.54&	1473-52908-0266&	4197-55479-0464&	    727.09&	         1&	    5151.6 -     131.8&	      1.51$\pm$0.48&	      0.10$\pm$0.03&	      0.86$\pm$0.08 \\ 
SDSS J215249.27$+$000722.8$^{\dagger}$&	      2.74&	1153-52933-0499&	4197-55479-0774&	    680.57&	         1&	    8032.5 -    4720.6&	      3.02$\pm$0.24&	      0.32$\pm$0.03&	      0.86$\pm$0.06 \\ 
SDSS J215631.24$+$003757.9 &	      1.86&	1107-52968-0629&	5147-55854-0202&	   1008.09&	         1&	   14180.2 -    3767.1&	     -0.57$\pm$1.30&	     -0.01$\pm$0.03&	      1.16$\pm$0.24 \\ 
SDSS J220328.52$+$005025.2  &	      2.55&	1474-52933-0617&	4198-55480-0954&	    717.46&	         1&	   11439.4 -    3575.4&	      3.04$\pm$1.11&	      0.13$\pm$0.05&	      1.02$\pm$0.16 \\ 
SDSS J220505.89$-$004722.4 &	      2.42&	1475-52903-0209&	4199-55481-0292&	    753.80&	         1&	    6606.2 -     422.4&	     -1.38$\pm$1.15&	     -0.06$\pm$0.05&	      1.19$\pm$0.20 \\ 
SDSS J220845.01$+$010007.6 &	      2.36&	1475-52903-0576&	4199-55481-0909&	    767.26&	         2&	   17782.4 -   12283.6&	     -4.40$\pm$0.95&	     -0.45$\pm$0.10&	      0.74$\pm$0.17 \\ 
               $ $         &	          &	 &	 &	          &	          &	    8118.2 -    4060.1&	     -5.03$\pm$0.72&	     -0.62$\pm$0.09&	      0.77$\pm$0.15 \\ 
SDSS J220850.31$-$004544.8 &	      1.69&	1105-52937-0055&	4200-55499-0449&	    954.19&	         1&	    5035.7 -     342.2&	     -2.03$\pm$1.42&	     -0.10$\pm$0.07&	      1.49$\pm$0.26 \\ 
SDSS J221242.48$-$004114.9$^{\dagger}$ &	      2.52&	1104-52912-0251&	4200-55499-0256&	    735.36&	         1&	    4001.1 -    1835.6&	      0.19$\pm$0.13&	      0.04$\pm$0.03&	      0.73$\pm$0.03 \\ 
SDSS J221326.95$+$003845.9$^{\dagger}$ &	      2.52&	1476-52964-0461&	4200-55499-0804&	    719.15&	         3&	   19468.7 -   17450.4&	      0.19$\pm$0.18&	      0.05$\pm$0.05&	      0.52$\pm$0.05 \\ 
               $ $         &	          &	 &	 &	          &	          &	   15131.5 -   12941.4&	      0.09$\pm$0.19&	      0.02$\pm$0.05&	      0.53$\pm$0.06 \\ 
               $ $         &	          &	 &	 &	          &	          &	   12383.2 -    6950.9&	      2.13$\pm$0.28&	      0.15$\pm$0.02&	      0.73$\pm$0.05 \\ 
SDSS J221555.99$+$010127.1$^{\dagger}$ &	      2.23&	1104-52912-0530&	4201-55443-0550&	    783.11&	         4&	   20378.9 -   18918.8&	     -0.16$\pm$0.11&	     -0.05$\pm$0.04&	      0.61$\pm$0.04 \\ 
               $ $         &	          &	 &	 &	          &	          &	   11210.6 -    8741.4&	     -1.10$\pm$0.15&	     -0.38$\pm$0.05&	      0.47$\pm$0.04 \\ 
               $ $         &	          &	 &	 &	          &	          &	    7216.9 -    4812.1&	     -0.80$\pm$0.14&	     -0.21$\pm$0.04&	      0.46$\pm$0.03 \\ 
               $ $         &	          &	 &	 &	          &	          &	    4125.0 -    3008.5&	     -0.10$\pm$0.09&	     -0.05$\pm$0.04&	      0.41$\pm$0.03 \\ 
SDSS J221558.15$-$005521.7$^{\dagger}$&	      2.55&	3146-53773-0013&	4201-55443-0458&	    470.56&	         2&	   17218.5 -   13087.4&	     -0.45$\pm$0.19&	     -0.05$\pm$0.02&	      0.69$\pm$0.03 \\ 
               $ $         &	          &	 &	 &	          &	          &	    8744.4 -    6467.7&	      0.11$\pm$0.15&	      0.03$\pm$0.03&	      0.62$\pm$0.04 \\ 
SDSS J221745.57$-$004925.4&	      3.14&	1476-52964-0019&	4201-55443-0418&	    599.23&	         1&	    4229.9 -    2150.2&	     -1.11$\pm$0.44&	     -0.23$\pm$0.09&	      0.63$\pm$0.12 \\ 
SDSS J222050.59$+$005948.6 &	      2.60&	1103-52873-0375&	4201-55443-0821&	    713.69&	         1&	   22772.1 -   10401.7&	      1.39$\pm$1.33&	      0.04$\pm$0.04&	      1.01$\pm$0.15 \\ 
SDSS J222106.05$-$005541.6 &	      2.32&	1103-52873-0208&	4201-55443-0177&	    774.10&	         1&	   19088.1 -   14615.4&	      5.73$\pm$0.74&	      0.97$\pm$0.12&	      0.51$\pm$0.15 \\ 
SDSS J222157.97$-$010331.0$^{\dagger}$ &	      2.67&	1137-52971-0123&	4201-55443-0116&	    672.84&	         2&	   17243.4 -   14754.0&	     -1.72$\pm$0.09&	     -0.40$\pm$0.02&	      0.60$\pm$0.02 \\ 
               $ $         &	          &	 &	 &	          &	          &	   12541.2 -    9599.6&	     -2.00$\pm$0.10&	     -0.79$\pm$0.04&	      0.45$\pm$0.02 \\ 
SDSS J222642.94$+$004535.6&	      3.28&	0673-52162-0372&	4202-55445-0798&	    766.34&	         2&	    6343.0 -    4260.3&	      0.03$\pm$0.41&	      0.01$\pm$0.14&	      0.53$\pm$0.11 \\ 
               $ $         &	          &	 &	 &	          &	          &	    2285.0 -     760.5&	     -0.52$\pm$0.32&	     -0.20$\pm$0.12&	      0.54$\pm$0.10 \\ 
SDSS J222834.61$-$011120.9 &	      2.38&	0673-52162-0123&	4202-55445-0043&	    972.45&	         1&	   18788.6 -   13902.2&	      2.57$\pm$0.61&	      0.51$\pm$0.12&	      0.45$\pm$0.11 \\ 
SDSS J223118.34$-$003321.4$^{\dagger}$ &	      2.53&	0673-52162-0101&	4202-55445-0032&	    930.03&	         1&	    8657.8 -    5565.9&	      1.80$\pm$0.25&	      0.50$\pm$0.07&	      0.45$\pm$0.06 \\ 
SDSS J223253.56$-$001119.4$^{\dagger}$ &	      3.09&	0673-52162-0029&	4203-55447-0240&	    803.18&	         2&	   19421.6 -   17083.4&	      0.66$\pm$0.61&	      0.12$\pm$0.11&	      0.74$\pm$0.16 \\ 
               $ $         &	          &	 &	 &	          &	          &	   12400.9 -    5573.3&	      4.59$\pm$0.78&	      0.28$\pm$0.05&	      0.88$\pm$0.13 \\ 
SDSS J223437.67$+$000326.5 &	      2.38&	1102-52883-0037&	4203-55447-0878&	    757.68&	         1&	    8412.6 -    5492.5&	     -1.72$\pm$1.81&	     -0.24$\pm$0.26&	      0.84$\pm$0.45 \\ 
SDSS J224206.88$+$002356.5 &	      2.42&	0675-52590-0400&	4205-55454-0529&	    837.43&	         1&	    8266.0 -    4057.0&	     -0.27$\pm$0.74&	     -0.02$\pm$0.05&	      1.10$\pm$0.15 \\ 
SDSS J224337.98$-$005408.1$^{\dagger}$ &	      1.57&	0675-52590-0201&	4205-55454-0408&	   1114.40&	         1&	    5028.0 -    286.8&	     -0.70$\pm$0.65&	     -0.04$\pm$0.03&	      1.02$\pm$0.10 \\ 
SDSS J224733.47$-$002009.1$^{\dagger}$&	      2.52&	0676-52178-0315&	4205-55454-0188&	    931.74&	         1&	    6625.9 -     690.5&	      1.62$\pm$0.51&	      0.07$\pm$0.02&	      0.96$\pm$0.09 \\ 
SDSS J225057.74$-$005630.5&	      3.13&	0379-51789-0286&	4206-55471-0380&	    890.45&	         2&	   14365.7 -    5885.3&	     -1.00$\pm$1.46&	     -0.03$\pm$0.05&	      1.22$\pm$0.22 \\ 
               $ $         &	          &	 &	 &	          &	          &	    2779.7 -     639.9&	      0.14$\pm$0.72&	      0.02$\pm$0.08&	      1.17$\pm$0.30 \\ 
SDSS J225608.48$+$010557.8$^{\dagger}$&	      2.27&	0676-52178-0604&	4206-55471-0905&	   1007.65&	         3&	   21591.9 -   17192.6&	      0.10$\pm$0.19&	      0.02$\pm$0.03&	      0.49$\pm$0.04 \\ 
               $ $         &	          &	 &	 &	          &	          &	    6510.8 -    3543.5&	     -0.01$\pm$0.11&	     -0.00$\pm$0.01&	      0.94$\pm$0.02 \\ 
               $ $         &	          &	 &	 &	          &	          &	    8386.9 -    6950.4&	      0.13$\pm$0.09&	      0.03$\pm$0.02&	      0.81$\pm$0.03 \\ 
SDSS J225706.16$-$002532.8$^{\dagger}$&	      1.99&	0380-51792-0274&	4206-55471-0091&	   1229.20&	         1&	    3077.3 -    305.0&	     -1.03$\pm$0.36&	     -0.07$\pm$0.02&	      1.09$\pm$0.07 \\ 
SDSS J225722.20$+$005033.2$^{\dagger}$&	      2.36&	0676-52178-0620&	4207-55475-0596&	    981.25&	         1&	   14272.0 -    4135.0&	     -1.28$\pm$0.90&	     -0.06$\pm$0.04&	      0.79$\pm$0.11 \\ 
SDSS J225752.55$+$002230.3$^{\dagger}$&	      3.27&	0677-52606-0391&	4207-55475-0604&	    671.74&	         1&	    8475.3 -    4887.6&	      0.63$\pm$0.30&	      0.10$\pm$0.05&	      0.62$\pm$0.06 \\

\hline
 \end{tabular} 
                \caption{Table continued on next page} 
                \label{d} 
         \end{table*} 

\setcounter{table}{0}
\begin{table*} 
         \centering 
         \fontsize{6}{9}\selectfont
         \begin{tabular}{lccccccccc} 
         \hline 
         SDSS name & z$^a$ & P-M-F$_1^b$ & P-M-F$_2^c$ &$\Delta$t$^{d}$ & n$_{tr}^e$ & $v_{max}$-$v_{min}^f$  & $\Delta$EW$^g$ &         $\Delta$EW/$\langle$EW$\rangle^h$ & $\langle d_{max}\rangle^i$  \\ 
        (J2000) & & & & (days)& &( km s$^{-1}$) &(\AA) & &  \\
          \hline 
SDSS J225819.00$-$000351.9&	      2.37&	0677-52606-0237&	4207-55475-0641&	    851.59&	         3&	   12385.4 -    7834.8&	      5.98$\pm$0.59&	      0.81$\pm$0.08&	      0.52$\pm$0.11 \\ 
               $ $         &	          &	 &	 &	          &	          &	    7558.9 -    3970.6&	      5.37$\pm$0.52&	      1.11$\pm$0.11&	      0.55$\pm$0.11 \\ 
               $ $         &	          &	 &	 &	          &	          &	    3768.8 -    1946.4&	      2.56$\pm$0.33&	      0.71$\pm$0.09&	      0.67$\pm$0.10 \\ 
SDSS J230639.64$+$010855.2$^{\dagger}$&	      3.65&	0678-52884-0401&	4208-55476-0748&	    557.69&	         1&	    5089.9 -    1375.3&	      0.75$\pm$0.21&	      0.12$\pm$0.03&	      0.54$\pm$0.04 \\ 
SDSS J231739.75$-$005719.4$^{\dagger}$&	      2.29&	0679-52177-0046&	4209-55478-0046&	   1003.04&	         1&	    9991.3 -    1780.5&	     10.78$\pm$0.43&	      0.47$\pm$0.02&	      0.94$\pm$0.05 \\ 
SDSS J231858.56$-$005049.6$^{\dagger}$&	      3.20&	0679-52177-0006&	4210-55444-0331&	    777.86&	         2&	    9730.2 -    7661.0&	     -1.70$\pm$0.35&	     -0.48$\pm$0.10&	      0.49$\pm$0.07 \\ 
               $ $         &	          &	 &	 &	          &	          &	    4141.9 -    1104.9&	     -0.70$\pm$0.20&	     -0.06$\pm$0.02&	      0.92$\pm$0.04 \\ 
SDSS J231911.92$-$001856.8$^{\dagger}$&	      2.31&	0679-52177-0079&	4210-55444-0314&	    987.01&	         1&	    4058.3 -    1642.5&	      2.57$\pm$0.46&	      0.43$\pm$0.08&	      0.75$\pm$0.12 \\ 
SDSS J232515.96$+$004649.8&	      3.09&	1485-52992-0459&	4211-55446-0680&	    599.56&	         1&	    3239.5 -     271.3&	     -2.53$\pm$1.07&	     -0.27$\pm$0.11&	      0.81$\pm$0.12 \\ 
SDSS J232859.37$-$000519.8&	      2.54&	0681-52199-0198&	4211-55446-0146&	    916.97&	         2&	   14813.7 -    6679.1&	     -2.44$\pm$1.09&	     -0.11$\pm$0.05&	      0.98$\pm$0.15 \\ 
               $ $         &	          &	 &	 &	          &	          &	    3021.5 -     536.6&	      0.21$\pm$0.33&	      0.03$\pm$0.04&	      0.83$\pm$0.06 \\ 
SDSS J232931.36$-$002036.2&	      3.00&	1485-52992-0021&	4212-55447-0466&	    613.14&	         1&	    7968.0 -    3413.7&	     -3.01$\pm$0.86&	     -0.23$\pm$0.06&	      0.92$\pm$0.15 \\ 
SDSS J232951.46$+$010706.6&	      1.73&	0681-52199-0374&	4211-55446-0906&	   1188.51&	         1&	   15608.8 -   10717.2&	      0.68$\pm$1.64&	      0.04$\pm$0.09&	      1.09$\pm$0.29 \\ 
SDSS J233024.66$-$003138.3&	      2.56&	0681-52199-0179&	4212-55447-0430&	    913.13&	         2&	   20906.4 -   17469.0&	      0.91$\pm$0.58&	      0.20$\pm$0.13&	      0.42$\pm$0.13 \\ 
               $ $         &	          &	 &	 &	          &	          &	   15611.0 -   12855.9&	     -0.86$\pm$0.51&	     -0.21$\pm$0.12&	      0.50$\pm$0.13 \\ 
SDSS J233845.19$-$000327.1&	      2.44&	4213-53449-0360&	8766-57333-0899&	    1129.88&	         1&	    3272.2 -     718.3&	      0.02$\pm$0.61&	      0.00$\pm$0.07&	      1.06$\pm$0.15 \\ 
SDSS J233939.85$+$001938.0$^{\dagger}$&	      2.84&	1093-52591-0422&	4213-55449-0732&	    743.69&	         1&	    5260.7 -    3190.2&	      1.21$\pm$1.04&	      0.56$\pm$0.48&	      0.35$\pm$0.16 \\ 
SDSS J234110.00$-$004159.1$^{\dagger}$ &	      3.81&	0682-52525-0017&	4213-55449-0215&	    607.90&	         1&	   18749.1 -    9451.5&	      0.79$\pm$0.70&	      0.03$\pm$0.03&	      0.99$\pm$0.09 \\ 
SDSS J234315.88$+$004659.5$^{\dagger}$ &	      2.78&	0385-51877-0454&	4213-55449-0918&	    944.66&	         1&	    3515.2 -    1030.2&	     -0.09$\pm$0.18&	     -0.01$\pm$0.02&	      0.97$\pm$0.04 \\ 
SDSS J234353.63$-$004450.7$^{\dagger}$ &	      3.22&	0683-52524-0212&	4213-55449-0082&	    692.85&	         2&	   15269.1 -   12307.2&	      0.91$\pm$0.23&	      0.16$\pm$0.04&	      0.63$\pm$0.05 \\ 
               $ $         &	          &	 &	 &	          &	          &	    9687.5 -    6238.6&	      1.13$\pm$0.21&	      0.18$\pm$0.03&	      0.51$\pm$0.05 \\ 
SDSS J234414.97$+$004855.5$^{\dagger}$&	      2.95&	0683-52524-0374&	4213-55449-0958&	    739.94&	         1&	    6885.7 -    1572.7&	     -0.13$\pm$0.48&	     -0.01$\pm$0.02&	      0.93$\pm$0.09 \\ 
SDSS J234847.18$+$002739.3$^{\dagger}$&	      2.56&	1903-53357-0629&	4214-55451-0857&	    589.03&	         3&	   15718.1 -   13376.7&	      2.61$\pm$0.31&	      1.02$\pm$0.12&	      0.33$\pm$0.08 \\ 
               $ $         &	          &	 &	 &	          &	          &	    3158.5 -    1419.3&	      1.48$\pm$0.24&	      0.29$\pm$0.05&	      0.77$\pm$0.07 \\ 
               $ $         &	          &	 &	 &	          &	          &	    5860.6 -    3859.2&	      1.68$\pm$0.26&	      0.22$\pm$0.03&	      0.97$\pm$0.07 \\ 
SDSS J234905.01$+$002855.0$^{\dagger}$ &	      2.35&	0684-52523-0396&	4215-55471-0526&	    878.69&	         1&	    2305.1 -   1189.6&	     -0.73$\pm$0.68&	     -0.08$\pm$0.08&	      0.79$\pm$0.13 \\ 
SDSS J235224.13$-$000951.0$^{\dagger}$ &	      2.74&	0684-52523-0196&	4215-55471-0350&	    787.79&	         2&	   20378.5 -   13428.8&	      2.58$\pm$0.61&	      0.22$\pm$0.05&	      0.61$\pm$0.09 \\ 
               $ $         &	          &	 &	 &	          &	          &	    7775.7 -    4808.7&	      2.50$\pm$0.41&	      0.35$\pm$0.06&	      0.86$\pm$0.10 \\ 
SDSS J235238.08$+$010552.3$^{\dagger}$ &	      2.15&	0386-51788-0524&	4215-55471-0676&	   1168.09&	         1&	    3434.4 -     121.0&	     -0.16$\pm$0.08&	     -0.01$\pm$0.01&	      0.93$\pm$0.01 \\ 
SDSS J235253.51$-$002850.4$^{\dagger}$&	      1.64&	0386-51788-0167&	4215-55471-0358&	   1395.60&	         1&	   17758.9 -   10757.6&	     -0.94$\pm$0.51&	     -0.04$\pm$0.02&	      0.95$\pm$0.07 \\ 
SDSS J235554.11$+$000444.5$^{\dagger}$&	      2.46&	0684-52523-0598&	4215-55471-0860&	    852.27&	         1&	   16138.7 -    6765.2&	     -6.11$\pm$1.10&	     -0.22$\pm$0.04&	      1.01$\pm$0.14 \\ 
SDSS J235610.98$-$001804.7&	      2.09&	4216-53477-0466&	8769-57338-0266&	    1249.27&	         1&	    3665.2 -     766.2&	     -0.16$\pm$1.00&	     -0.01$\pm$0.09&	      1.14$\pm$0.24 \\ 
SDSS J235628.96$-$003602.0$^{\dagger}$ &	      2.95&	0387-51791-0315&	4215-55471-0137&	    931.61&	         3&	    7494.7 -    4734.7&	     -2.14$\pm$0.21&	     -0.45$\pm$0.04&	      0.52$\pm$0.05 \\ 
               $ $         &	          &	 &	 &	          &	          &	    4044.5 -    2042.9&	     -1.94$\pm$0.15&	     -0.90$\pm$0.07&	      0.38$\pm$0.04 \\ 
               $ $         &	          &	 &	 &	          &	          &	    1286.7 -    -214.1&	      0.19$\pm$0.08&	      0.04$\pm$0.02&	      0.77$\pm$0.02 \\ 
SDSS J235702.54$-$004824.0$^{\dagger}$ &	      3.00&	0387-51791-0246&	4216-55477-0410&	    921.96&	         2&	    9518.9 -    7518.7&	     -1.36$\pm$0.19&	     -0.52$\pm$0.07&	      0.38$\pm$0.05 \\ 
               $ $         &	          &	 &	 &	          &	          &	    3516.4 -     479.2&	     -0.93$\pm$0.14&	     -0.08$\pm$0.01&	      0.99$\pm$0.03 \\ 
SDSS J235708.69$+$003929.6$^{\dagger}$&	      2.49&	1091-52902-0405&	4215-55471-0929&	    735.13&	         1&	   16437.5 -   13157.2&	      0.39$\pm$0.47&	      0.12$\pm$0.15&	      0.34$\pm$0.11 \\ 
SDSS J235730.49$+$002538.9$^{\dagger}$&	      2.51&	0684-52523-0632&	4215-55471-0924&	    839.89&	         1&	    1543.2 -     321.1&	      0.30$\pm$0.29&	      0.19$\pm$0.18&	      0.34$\pm$0.09 \\ 
SDSS J235859.47$-$002426.2$^{\dagger}$&	      1.76&	0387-51791-0181&	4216-55477-0394&	   1336.96&	         1&	   15234.2 -    3443.2&	      4.67$\pm$0.42&	      0.23$\pm$0.02&	      0.90$\pm$0.04 \\ 
\hline
 \end{tabular} 
 \begin{flushleft}
 $^\dagger$ The sources in the high-SNR sample. \\
 $^a$ visual inspection redshift from DR12Q catalog \\
 $^b$ Plate-MJD-Fiber information of the the first epoch\\ 
 $^c$ Plate-MJD-Fiber information of the the second epoch\\
 $^d$ Rest-frame time-scale difference between spectroscopic observations\\
 $^e$ Number of absorption troughs\\
 $^f$ minimum and maximum velocity of absorption troughs\\
 $^g$ Rest-frame equivalent width difference between epoch 1 and epoch 2 spectra \\
 $^h$ Rest-frame fractional equivalent width difference between epoch 1 and epoch 2 spectra\\
 $^i$ Average maximum trough depth\\
 \end{flushleft}
                \caption{Table listing the measurements of absorption line variability parameters for sources in the final sample.} 
                \label{appendix_table1} 
         \end{table*}

\begin{table*}
         \centering 
        \fontsize{7}{9}\selectfont
         \begin{tabular}{lccccccccccc} 
         \hline 
         SDSS name & VAR\_& VAR\_& VAR\_& n$_{epoch}^d$ & SF$_\infty^e$ & $\sigma$(m)$^f$ & $\langle\Delta$m$\rangle^g$ &$\langle\Delta$m/$\Delta$t$\rangle^h$ & $\sigma$(m)$^i$ & $\langle\Delta$m$\rangle^j$ & $\langle\Delta$m/$\Delta$t$\rangle^k$ \\ 
         
(J2000)& A$^a$ & GAMMA$^b$ & CHI2$^c$ & & & SDSS & SDSS & SDSS& CRTS& CRTS& CRTS \\               
 \hline 
SDSS J210946.54$-$003633.0  &	      0.09&	      0.53&	     21.92&	      47.0&	    -&	      0.18&	      0.20&	      0.44&	      0.11&	      0.12&	      0.14	\\
SDSS J211200.99$-$003443.5  &	      0.07&	      0.42&	     16.73&	      67.0&	    -&	      0.05&	      0.05&	      0.11&	      -&	      -&	      -	\\
SDSS J211415.82$-$011316.5  &	      0.03&	      0.59&	      3.43&	      38.0&	      0.20$^{  0.91}_{  0.18}$&	      0.07&	      0.06&	      0.18&	      0.30&	      0.31&	      0.34	\\
SDSS J211518.49$+$001115.2  &	      0.05&	      0.26&	      4.18&	      57.0&	    -&	      0.14&	      0.14&	      0.36&	      -&	      -&	      -	\\
SDSS J211829.90$+$002830.3  &	      0.04&	      0.47&	     11.49&	      69.0&	    -&	      0.14&	      0.16&	      0.28&	      0.20&	      0.23&	      0.25	\\
SDSS J211853.67$+$002843.9$^{\dagger}$&	      0.03&	      0.25&	      3.26&	      57.0&	      0.04$^{  0.29}_{  0.03}$&	      0.04&	      0.04&	      0.12&	      0.06&	      0.07&	      0.06	\\
SDSS J211908.43$+$003246.3$^{\dagger}$&	      0.07&	      0.20&	     22.24&	      59.0&	      0.12$^{  0.95}_{  0.05}$&	      0.06&	      0.06&	      0.22&	      0.08&	      0.09&	      0.11	\\
SDSS J212026.14$+$000735.0  &	      0.03&	      0.61&	      6.51&	      52.0&	    -&	      0.11&	      0.11&	      0.22&	      0.16&	      0.19&	      0.19	\\
SDSS J212204.46$-$001012.5  &	      0.07&	      0.34&	      6.24&	      45.0&	      0.17$^{  0.48}_{  0.11}$&	      0.12&	      0.12&	      0.37&	      0.02&	      0.03&	      0.03	\\
SDSS J212319.66$-$002245.8$^{\dagger}$&	      0.04&	      0.57&	     12.43&	      61.0&	    -&	      0.08&	      0.08&	      0.29&	      0.34&	      0.69&	      0.40	\\
SDSS J212419.26$+$004313.4$^{\dagger}$&	      0.06&	      0.34&	     13.75&	      49.0&	      0.08$^{  0.73}_{  0.05}$&	      0.06&	      0.06&	      0.13&	      0.12&	      0.13&	      0.09	\\
SDSS J212811.62$-$010704.8$^{\dagger}$&	      0.06&	      0.24&	      6.16&	      58.0&	    -&	      0.08&	      0.09&	      0.45&	      0.13&	      0.15&	      0.23	\\
SDSS J212952.14$+$003244.5$^{\dagger}$&	      0.05&	      0.33&	      8.83&	      57.0&	      0.13$^{  0.35}_{  0.09}$&	      0.09&	      0.07&	      0.18&	      0.07&	      0.07&	      0.06	\\
SDSS J213008.34$+$002100.0$^{\dagger}$&	      0.07&	      0.49&	     17.31&	      77.0&	      0.20$^{  1.17}_{  0.11}$&	      0.12&	      0.12&	      0.39&	      0.08&	      0.09&	      0.09	\\
SDSS J213138.07$-$002537.8$^{\dagger}$&	      0.03&	      0.73&	     18.66&	      71.0&	    -&	      0.07&	      0.05&	      0.12&	      0.07&	      0.07&	      0.08	\\
SDSS J213648.17$-$001546.6$^{\dagger}$&	      0.07&	      0.36&	     20.43&	      55.0&	    -&	      0.06&	      0.04&	      0.18&	      0.13&	      0.15&	      0.17	\\
SDSS J213901.65$-$004925.2  &	      0.04&	      0.41&	      4.39&	      48.0&	      0.13$^{  1.11}_{  0.08}$&	      0.09&	      0.08&	      0.40&	      0.29&	      0.33&	      0.47	\\
SDSS J214107.19$+$001213.4  &	      0.02&	      0.46&	      1.19&	      81.0&	      0.05$^{  0.22}_{  0.05}$&	      0.05&	      0.05&	      0.22&	      0.53&	      0.39&	      0.93	\\
SDSS J214113.05$-$003545.9$^{\dagger}$&	      0.04&	      0.19&	      5.60&	      58.0&	      0.07$^{  0.55}_{  0.04}$&	      0.04&	      0.04&	      0.10&	      0.08&	      0.09&	      0.11	\\
SDSS J214227.48$+$005651.7$^{\dagger}$&	      0.05&	      0.38&	      3.31&	      59.0&	      0.25$^{  1.63}_{  0.14}$&	      0.10&	      0.09&	      0.45&	      0.72&	      0.60&	      0.81	\\
SDSS J214456.61$+$005926.2  &	      0.16&	      0.44&	     19.90&	      57.0&	      0.34$^{  2.43}_{  0.17}$&	      0.23&	      0.22&	      0.52&	      0.72&	      0.60&	      0.81	\\
SDSS J214701.79$+$001643.3  &	      0.04&	      0.58&	      8.30&	      68.0&	      0.21$^{  1.39}_{  0.12}$&	      0.09&	      0.09&	      0.18&	      0.19&	      0.21&	      0.21	\\
SDSS J214821.43$-$003705.8  &	      0.03&	      0.69&	      7.14&	      53.0&	    -&	      0.09&	      0.09&	      0.21&	      0.20&	      0.26&	      0.32	\\
SDSS J215249.27$+$000722.8$^{\dagger}$&	      0.08&	      0.22&	     10.32&	      58.0&	      0.14$^{  0.97}_{  0.08}$&	      0.09&	      0.09&	      0.33&	      0.11&	      0.14&	      0.15	\\
SDSS J215631.24$+$003757.9  &	      0.07&	      0.20&	      5.21&	     124.0&	      0.12$^{  0.50}_{  0.09}$&	      0.10&	      0.09&	      0.22&	      0.19&	      0.21&	      0.21	\\
SDSS J220328.52$+$005025.2  &	      0.10&	      0.45&	     12.73&	     125.0&	      0.28$^{  1.15}_{  0.18}$&	      0.17&	      0.13&	      0.49&	      0.24&	      0.27&	      0.33	\\
SDSS J220505.89$-$004722.4  &	      0.05&	      0.26&	      2.59&	      54.0&	      0.18$^{  1.39}_{  0.11}$&	      0.13&	      0.12&	      0.39&	      0.29&	      0.38&	      0.32	\\
SDSS J220845.01$+$010007.6  &	      0.08&	      0.18&	      3.98&	      44.0&	      0.11$^{  0.52}_{  0.08}$&	      0.11&	      0.09&	      0.28&	      0.25&	      0.22&	      0.37	\\
SDSS J220850.31$-$004544.8  &	      0.07&	      0.59&	      9.33&	      52.0&	    -&	      0.08&	      0.09&	      0.19&	      0.54&	      0.58&	      0.49	\\
SDSS J221242.48$-$004114.9$^{\dagger}$&	      0.05&	      0.29&	      6.36&	      45.0&	      0.12$^{  0.79}_{  0.06}$&	      0.06&	      0.06&	      0.22&	      0.10&	      0.12&	      0.14	\\
SDSS J221326.95$+$003845.9$^{\dagger}$&	      0.05&	      0.54&	     24.52&	      71.0&	    -&	      0.10&	      0.09&	      0.25&	      0.26&	      0.35&	      0.31	\\
SDSS J221555.99$+$010127.1$^{\dagger}$&	      0.04&	      0.46&	     11.07&	      69.0&	    -&	      0.05&	      0.04&	      0.11&	      0.09&	      0.10&	      0.11	\\
SDSS J221558.15$-$005521.7$^{\dagger}$&	      0.03&	      0.45&	      6.72&	      59.0&	    -&	      0.07&	      0.06&	      0.13&	      0.06&	      0.06&	      0.07	\\
SDSS J221745.57$-$004925.4  &	      0.06&	      0.33&	      5.53&	      52.0&	      0.17$^{  1.15}_{  0.09}$&	      0.12&	      0.13&	      0.56&	      0.33&	      0.40&	      0.64	\\
SDSS J222050.59$+$005948.6  &	      0.02&	      0.66&	      5.54&	      70.0&	      0.16$^{  0.80}_{  0.11}$&	      0.07&	      0.04&	      0.16&	      0.10&	      0.11&	      0.13	\\
SDSS J222106.05$-$005541.6  &	      0.05&	      0.69&	     18.03&	      35.0&	    -&	      0.13&	      0.11&	      0.28&	      0.25&	      0.28&	      0.31	\\
SDSS J222157.97$-$010331.0$^{\dagger}$&	      0.02&	      0.50&	      2.70&	      48.0&	      0.05$^{  0.20}_{  0.03}$&	      0.03&	      0.03&	      0.10&	      0.12&	      0.14&	      0.13	\\
SDSS J222642.94$+$004535.6  &	      0.06&	      0.56&	     13.95&	      63.0&	      0.27$^{  1.73}_{  0.12}$&	      0.14&	      0.11&	      0.36&	      0.28&	      0.20&	      0.31	\\
SDSS J222834.61$-$011120.9  &	      0.02&	      0.79&	     10.08&	      51.0&	    -&	      0.11&	      0.07&	      0.23&	      0.21&	      0.20&	      0.27	\\
SDSS J223118.34$-$003321.4$^{\dagger}$&	      0.05&	      0.50&	     18.31&	      54.0&	    -&	      0.10&	      0.09&	      0.24&	      0.16&	      0.17&	      0.20	\\
SDSS J223253.56$-$001119.4$^{\dagger}$&	      0.05&	      0.29&	      4.18&	      56.0&	      0.12$^{  0.89}_{  0.07}$&	      0.08&	      0.09&	      0.24&	      0.42&	      0.44&	      0.55	\\
SDSS J223437.67$+$000326.5  &	      0.09&	      0.29&	      7.72&	      57.0&	    -&	      0.08&	      0.09&	      0.20&	      0.33&	      0.37&	      0.41	\\
SDSS J224206.88$+$002356.5  &	      0.12&	      0.35&	     24.15&	      57.0&	      0.39$^{  2.03}_{  0.23}$&	      0.16&	      0.16&	      0.38&	      0.19&	      0.20&	      0.26	\\
SDSS J224337.98$-$005408.1$^{\dagger}$&	      0.08&	      0.36&	     11.56&	      57.0&	    -&	      0.11&	      0.09&	      0.15&	      0.16&	      0.16&	      0.16	\\
SDSS J224733.47$-$002009.1$^{\dagger}$&	      0.04&	      0.23&	      3.78&	      57.0&	      0.09$^{  0.56}_{  0.06}$&	      0.05&	      0.04&	      0.15&	      0.15&	      0.16&	      0.18	\\
SDSS J225057.74$-$005630.5  &	      0.01&	      0.85&	      1.75&	      42.0&	      0.09$^{  0.27}_{  0.07}$&	      0.09&	      0.07&	      0.20&	      0.31&	      0.29&	      0.42	\\
SDSS J225608.48$+$010557.8$^{\dagger}$&	      0.05&	      0.48&	     19.33&	      75.0&	    -&	      0.07&	      0.07&	      0.15&	      0.10&	      0.11&	      0.15	\\
SDSS J225706.16$-$002532.8$^{\dagger}$&	      0.04&	      0.46&	      6.41&	      75.0&	      0.13$^{  0.70}_{  0.08}$&	      0.05&	      0.05&	      0.10&	      0.09&	      0.10&	      0.11	\\
SDSS J225722.20$+$005033.2$^{\dagger}$&	      0.04&	      0.21&	      3.95&	     115.0&	      0.08$^{  0.47}_{  0.05}$&	      0.05&	      0.04&	      0.13&	      0.06&	      0.07&	      0.10	\\
SDSS J225752.55$+$002230.3$^{\dagger}$&	      0.02&	      0.26&	      2.45&	      67.0&	      0.08$^{  0.52}_{  0.06}$&	      0.05&	      0.04&	      0.10&	      0.08&	      0.11&	      0.16	\\
SDSS J225819.00$-$000351.9  &	      0.06&	      0.28&	      5.67&	      63.0&	      0.14$^{  0.75}_{  0.09}$&	      0.07&	      0.07&	      0.15&	      0.18&	      0.18&	      0.23	\\
SDSS J230639.64$+$010855.2$^{\dagger}$&	      0.03&	      0.50&	      5.97&	      70.0&	    -&	      0.06&	      0.06&	      0.18&	      0.12&	      0.14&	      0.21	\\
SDSS J231739.75$-$005719.4$^{\dagger}$&	      0.05&	      0.72&	     34.97&	      67.0&	    -&	      0.15&	      0.14&	      0.27&	      0.18&	      0.20&	      0.30	\\
SDSS J231858.56$-$005049.6$^{\dagger}$&	      0.02&	      0.84&	     15.63&	      36.0&	    -&	      0.17&	      0.14&	      0.28&	      0.12&	      0.15&	      0.26	\\
SDSS J231911.92$-$001856.8$^{\dagger}$&	      0.08&	      0.34&	     12.95&	      54.0&	      0.20$^{  0.59}_{  0.13}$&	      0.15&	      0.13&	      0.32&	      0.12&	      0.14&	      0.21	\\
SDSS J232515.96$+$004649.8  &	      0.10&	      0.49&	     27.46&	      68.0&	      0.37$^{  1.90}_{  0.24}$&	      0.19&	      0.19&	      0.65&	      0.34&	      0.40&	      0.47	\\
SDSS J232859.37$-$000519.8  &	      0.01&	      1.02&	      3.30&	      61.0&	    -&	      0.07&	      0.06&	      0.13&	      0.08&	      0.09&	      0.11	\\
SDSS J232931.36$-$002036.2  &	      0.06&	      0.31&	      6.02&	      46.0&	      0.20$^{  1.28}_{  0.10}$&	      0.11&	      0.12&	      0.37&	      0.11&	      0.12&	      0.14	\\
SDSS J232951.46$+$010706.6  &	      0.06&	      0.83&	     32.15&	      70.0&	    -&	      0.74$^*$&	      0.14&	      0.42&	      0.14&	      0.17&	      0.15	\\
SDSS J233024.66$-$003138.3  &	      0.02&	      0.81&	     15.04&	      75.0&	    -&	      0.11&	      0.08&	      0.18&	      0.16&	      0.20&	      0.23	\\
SDSS J233845.19$-$000327.1  &	      0.02&	      0.87&	      2.79&	      50.0&	    -&	      0.14&	      0.14&	      0.32&	      0.06&	      0.06&	      0.07	\\
SDSS J233939.85$+$001938.0$^{\dagger}$&	      0.03&	      0.83&	     25.20&	      62.0&	    -&	      0.14&	      0.14&	      0.32&	      0.12&	      0.12&	      0.14	\\
SDSS J234110.00$-$004159.1$^{\dagger}$&	      0.00&	      1.26&	      6.39&	      52.0&	    -&	      0.09&	      0.07&	      0.26&	      0.12&	      0.13&	      0.26	\\
SDSS J234315.88$+$004659.5$^{\dagger}$&	      0.04&	      0.55&	     12.51&	      63.0&	    -&	      0.09&	      0.08&	      0.19&	      0.21&	      0.25&	      0.25	\\
SDSS J234353.63$-$004450.7$^{\dagger}$&	      0.06&	      0.54&	     33.00&	      53.0&	      0.39$^{  1.44}_{  0.29}$&	      0.12&	      0.11&	      0.43&	      0.15&	      0.18&	      0.27	\\
SDSS J234414.97$+$004855.5$^{\dagger}$&	      0.04&	      0.72&	     30.51&	      77.0&	    -&	      0.17&	      0.18&	      0.36&	      0.08&	      0.10&	      0.14	\\
SDSS J234847.18$+$002739.3$^{\dagger}$&	      0.05&	      0.56&	     13.98&	      62.0&	      0.16$^{  0.98}_{  0.08}$&	      0.10&	      0.09&	      0.28&	      0.07&	      0.08&	      0.10	\\

\hline

\end{tabular} 
                \caption{Table continued on next page.} 
                \label{a1} 
         \end{table*} 
         \setcounter{table}{1}
\begin{table*}
         \centering 
        \fontsize{7}{9}\selectfont
         \begin{tabular}{lccccccccccc} 
         \hline 
        SDSS name & VAR\_& VAR\_& VAR\_& n$_{epoch}^d$ & SF$_\infty^e$ & $\sigma$(m)$^f$ & $\langle\Delta$m$\rangle^g$ &$\langle\Delta$m/$\Delta$t$\rangle^h$ & $\sigma$(m)$^i$ & $\langle\Delta$m$\rangle^j$ & $\langle\Delta$m/$\Delta$t$\rangle^k$ \\ 
         (J2000)& A$^a$ & GAMMA$^b$ & CHI2$^c$ & & & SDSS & SDSS & SDSS& CRTS& CRTS& CRTS \\               

         \hline
SDSS J234905.01$+$002855.0$^{\dagger}$&	      0.06&	      0.33&	      4.52&	      66.0&	      0.17$^{  1.27}_{  0.15}$&	      0.08&	      0.06&	      0.17&	      0.15&	      0.11&	      0.17	\\
SDSS J235224.13$-$000951.0$^{\dagger}$&	      0.06&	      0.25&	      6.68&	      75.0&	      0.08$^{  0.60}_{  0.04}$&	      0.06&	      0.05&	      0.17&	      0.07&	      0.09&	      0.11	\\
SDSS J235238.08$+$010552.3$^{\dagger}$&	      0.05&	      0.18&	      7.59&	      28.0&	      0.15$^{  0.72}_{  0.10}$&	      0.06&	      0.05&	      0.11&	      0.09&	      0.04&	      0.06	\\
SDSS J235253.51$-$002850.4$^{\dagger}$&	      0.03&	      0.60&	     10.03&	      71.0&	    -&	      0.06&	      0.06&	      0.09&	      0.09&	      0.10&	      0.09	\\
SDSS J235554.11$+$000444.5$^{\dagger}$&	      0.09&	      0.35&	     12.25&	      54.0&	      0.28$^{  1.51}_{  0.17}$&	      0.11&	      0.11&	      0.38&	      0.09&	      0.10&	      0.09	\\
SDSS J235610.98$-$001804.7  &	      0.00&	      1.73&	      1.29&	      32.0&	    -&	      0.10&	      0.11&	      0.18&	      -&	      -&	      -	\\
SDSS J235628.96$-$003602.0$^{\dagger}$&	      0.02&	      0.32&	      2.93&	      62.0&	      0.07$^{  0.45}_{  0.04}$&	      0.05&	      0.03&	      0.09&	      0.13&	      0.14&	      0.24	\\
SDSS J235702.54$-$004824.0$^{\dagger}$&	      0.01&	      0.97&	      5.95&	      48.0&	      0.14$^{  1.10}_{  0.08}$&	      0.05&	      0.04&	      0.13&	      -&	      -&	      -	\\
SDSS J235708.69$+$003929.6$^{\dagger}$&	      0.03&	      0.67&	      8.10&	      75.0&	      0.18$^{  0.98}_{  0.10}$&	      0.08&	      0.05&	      0.14&	      0.30&	      0.43&	      0.38	\\
SDSS J235730.49$+$002538.9$^{\dagger}$&	      0.04&	      0.35&	      3.77&	      64.0&	      0.14$^{  0.71}_{  0.10}$&	      0.07&	      0.07&	      0.19&	      0.20&	      0.26&	      0.26	\\
SDSS J235859.47$-$002426.2$^{\dagger}$&	      0.06&	      0.47&	     23.28&	      50.0&	      0.31$^{  1.09}_{  0.25}$&	      0.08&	      0.09&	      0.19&	      0.08&	      0.09&	      0.08	\\

\hline
 \end{tabular}
 %\begin{tablenotes}
 \begin{flushleft}
 $^{\dagger}$ The sources in the high-SNR sample. \\
 $^*$ Note that the $\sigma$(m)$_{SDSS}$ value for the source SDSS J232951.46$+$010706.6  is unusually high as compared to other sources. All the other continuum variability parameters are relatively higher for this source. In all the five bands of SDSS light curves, there are three photometric observations ($\sim$ MJD 53000) during which the source weakened by two magnitudes. After a short period, the source brightness returned the value before MJD 53000. As the source is not part of the high-SNR sample, we included this source in our correlation analysis, but excluded it from the $\sigma$(m)$_{SDSS}$ vs. absorption line variability parameters in Fig.~\ref{fig:correlation_plot}. \\
 $^a$ VAR\_A parameter from DR12Q catalog (in magnitudes/year).  \\
 $^b$ VAR\_GAMMA parameter from DR12Q catalog.   \\
 $^c$ VAR\_CHI2 parameter from DR12Q catalog.  \\
 $^d$ Number of photometric epochs in the SDSS light curve.  \\
 $^e$ SF$_\infty$ parameter from CM10 catalog (in magnitudes). \\
 $^f$ Standard deviation of magnitudes in the SDSS light curve (in magnitudes).  \\
 $^g$ Median magnitude differences in the SDSS light curve (in magnitudes).  \\
 $^h$ Median rate of change of magnitude differences in the SDSS light curve (in magnitudes/year).  \\
 $^i$  Standard deviation of magnitudes in the CRTS light curve (in magnitudes). \\
 $^j$  Median magnitude differences in the CRTS light curve (in magnitudes). \\
 $^k$  Median rate of change of magnitude differences in the CRTS light curve (in magnitudes/year. \\
 \end{flushleft}

 %\end{tablenotes}
 
                \caption{Table listing the measurements of continuum variability parameters for sources in the final sample.} 
                \label{appendix_table2} 
         \end{table*} 

\begin{landscape}
\begin{table}
    \centering
    \fontsize{10}{10}\selectfont
    \begin{tabular}{lcccccccccc}%{ |p{3cm} | *{10}{c} }
    \hline
    \hline
                                            &  VAR\_A    & VAR\_GAMMA  &VAR\_CHI2 & SF$_{\infty}$ & $\sigma$(m) & $\langle\Delta$m$\rangle$ &  $\langle\Delta$m/$\Delta$t$\rangle$ & $\sigma$(m) & $\langle\Delta$m$\rangle$ &$\langle\Delta$m/$\Delta$t$\rangle$ \\
                                           &       &           &           &           &SDSS       &    SDSS    &  SDSS    &    CRTS   &  CRTS  & CRTS\\
        \hline
        VAR\_A                                   & 1.00 (0.00)  &           &           &           &       &        &      &       &       &\\
        VAR\_GAMMA                            &-0.15 (0.19)  & 1.00 (0.00)&           &           &       &        &      &       &    &\\
        VAR\_CHI2                           & 0.67 (0.00)  & 0.27 (0.02)& 1.00 (0.00) &           &       &        &      &       &    &\\
        SF$_{\infty}$                    & 0.72 (0.00)  & 0.44 (0.00)& 0.61 (0.00) & 1.00 (0.00)&       &        &      &       &    &\\
        $\sigma(m)_{SDSS}$                  & 0.63 (0.00)  & 0.30 (0.01)& 0.44 (0.00) & 0.79 (0.00)& 1.00 (0.00)&        &      &       &    &\\
        $\langle\Delta$m$\rangle_{SDSS}$             & 0.68 (0.00)  & 0.21 (0.06)& 0.42 (0.00) & 0.76 (0.00)& 0.95 (0.00)& 1.00 (0.00)&      &       &    &\\
        $\langle\Delta$m/$\Delta$t$\rangle_{SDSS}$   & 0.66 (0.00)  & 0.06 (0.59)& 0.31 (0.01) & 0.67 (0.00)& 0.83 (0.00)& 0.85 (0.00)& 1.00 (0.00)   &       &    &\\
        $\sigma(m)_{CRTS}$                  & 0.24 (0.04)  & 0.06 (0.55)& -0.03 (0.74)& 0.28 (0.06)& 0.30 (0.01)& 0.29 (0.01)& 0.38 (0.00)& 1.00 (0.00)  &    &\\
        $\langle\Delta$m$\rangle_{CRTS}$             & 0.25 (0.03)  & 0.08 (0.47)& -0.01 (0.92)& 0.29 (0.05)& 0.33 (0.00)& 0.33 (0.00)& 0.42 (0.00)& 0.98 (0.00)& 1.00 (0.00) &\\
        $\langle\Delta$m/$\Delta$t$\rangle_{CRTS}$   & 0.19 (0.08)  & 0.09 (0.46)& -0.06 (0.62)& 0.39 (0.01)& 0.30 (0.01)& 0.29 (0.01)& 0.43 (0.00)& 0.94 (0.00)& 0.95 (0.00)    &1.00(0.00)\\
        \hline
    \end{tabular}
    \caption{Spearman's Rank Correlation analysis of light curve variability parameters  between themselves.}
    \label{tab:appendix_correlationtable}
\end{table}
\end{landscape}
%%%%%%%%%%%%%%%%%%%%%%%%%%%%%%%%%%%%%%%%%%%%%%%%%%

% Don't change these lines
\bsp	% typesetting comment
\label{lastpage}
\end{document}